\documentclass[pre,twocolumn,showpacs]{revtex4}

\begin{document}
%%%%%%%%%%%
%%% MACROS
%%%%%%%%%%%

\def\kket{\rangle \mskip -3mu \rangle}
\def\bbra{\langle \mskip -3mu \langle}

\def\ket{\right\rangle}
\def\bra{\left\langle}
\def\kket{\rangle\!\rangle}
\def\bbra{\langle\!\langle}
\def\ckket{)\! )}
\def\cbbra{(\!(}

\def\pard{\partial}

\def\sinh{{\rm sinh}}
\def\sgn{{\rm sgn}}

%Characters
\def\alp{\alpha}
\def\del{\delta}
\def\Del{\Delta}
\def\eps{\epsilon}
\def\gam{\gamma}
\def\sig{\sigma}
\def\kap{\kappa}
\def\lam{\lambda}
\def\Lam{\Lambda}
\def\ome{\omega}
\def\Ome{\Omega}
\def\c{c}
\def\d{d}
\def\m{m}
\def\n{n}
\def\q{q}
\def\t{t}
\def\th{\theta}
\def\tht{{\tilde{\theta}}}
\def\vphi{\varphi}

\def\Gam{\Gamma}

\def\kav{{\bar k}}

\def\abf{{\bf a}}
\def\cbf{{\bf c}}
\def\dbf{{\bf d}}
\def\gbf{{\bf g}}
\def\kbf{{\bf k}}
\def\lbf{{\bf l}}
\def\nbf{{\bf n}}
\def\pbf{{\bf p}}
\def\qbf{{\bf q}}
\def\rbf{{\bf r}}
\def\ubf{{\bf u}}
\def\vbf{{\bf v}}
\def\xbf{{\bf x}}
\def\Cbf{{\bf C}}
\def\Dbf{{\bf D}}
\def\Kbf{{\bf K}}
\def\Pbf{{\bf P}}
\def\Qbf{{\bf Q}}

\def\omet{{\tilde \ome}}
\def\thetat{{\tilde \theta}}
\def\Qt{{\tilde Q}}
\def\Ft{{\tilde F}}
\def\At{{\tilde A}}
\def\Bt{{\tilde B}}
\def\Ht{{\tilde H}}
\def\mt{{\tilde m}}
\def\ut{{\tilde u}}
\def\bt{{\tilde b}}
\def\vt{{\tilde v}}
\def\nt{{\tilde n}}
\def\xt{{\tilde x}}
\def\wt{{\tilde w}}
\def\phit{{\tilde \phi}}
\def\rhot{{\tilde \rho}}
\def\alpt{{\tilde \alp}}
\def\Jt{{\tilde{J}}}

\def\Jv{{\vec J}}
\def\alv{{\vec \alpha}}

\def\Cb{{\bar C}}
\def\Ab{{\bar A}}
\def\Bb{{\bar B}}
\def\Db{{\bar D}}
\def\Nb{{\bar N}}
\def\Qb{{\bar Q}}
\def\Fb{{\bar F}}
\def\Vb{{\bar V}}
\def\gb{{\bar g}}
\def\nb{{\bar n}}
\def\bb{{\bar b}}
\def\cb{{\bar c}}
\def\qb{{a}}
\def\vb{{\bar v}}
\def\wb{{\bar w}}
\def\Pib{{\bar \Pi}}
\def\xib{{\bar \xi}}
\def\rhob{{\bar \rho}}
\def\phib{{\bar \phi}}
\def\chib{{\bar \chi}}
\def\psib{{\bar \psi}}
\def\omeb{{\bar \ome}}

\def\Sh{{\hat S}}
\def\Wh{{\hat W}}
\def\hz{{\hat z}}
\def\hq{{\hat q}}
\def\home{{\hat \omega}}
\def\hgam{{\hat \gamma}}
\def\hGam{{\hat \Gamma}}
\def\hrho{{\hat \rho}}
\def\hx{{\hat x}}
\def\hp{{\hat p}}
\def\hR{{\hat R}}

\def\psiw{{\xi}}
\def\tI{{g}}
\def\ad{a^\dagger}
\def\Ad{A^\dagger}
\def\Lamd{\Lam^{\dagger}}

\def\Ep#1{Eq.\ (\ref{#1})}
\def\Eqs#1{Eqs.\ (\ref{#1})}
\def\EQN#1{\label{#1}}
\def\comb#1#2{\left(\begin{array}{c}{#1} \\ {#2} \end{array}\right)}

\newcommand{\beqa}{\begin{eqnarray}}
\newcommand{\eeqa}{\end{eqnarray}}

%%%%%%%%%%%%%%%%%%%%%
% Title and authors %
%%%%%%%%%%%%%%%%%%%%%

\title{Star-unitary transformations. \\ From dynamics to irreversibility and stochastic behavior}
\author{Sungyun Kim}
\affiliation{Center for Studies in Statistical Mechanics and Complex Systems,
The University of Texas at Austin, Austin, TX 78712 USA}
\author{Gonzalo Ordonez }
\affiliation{Center for Studies in Statistical Mechanics and Complex Systems,
The University of Texas at Austin, Austin, TX 78712 USA\\ and International Solvay Institutes for Physics and Chemistry,
             CP231, 1050 Brussels, Belgium}
\date{\today}
\pacs{02.50.Fz, 05.40.-a, 05.70.Ln}

%%%%%%%%%%%%%%%%%%%%
% Abstract         %
%%%%%%%%%%%%%%%%%%%%

\begin{abstract}
 We consider a simple model of a classical harmonic oscillator coupled to a field.  In standard approaches Langevin-type equations for {\it bare} particles are derived from Hamiltonian dynamics. These equations contain memory terms and are time-reversal invariant. In contrast the phenomenological Langevin equations have no memory terms (they are Markovian equations) and give a time evolution split in two branches (semigroups), each of which breaks time symmetry.  A standard approach to bridge dynamics with phenomenology is to consider the Markovian approximation of the former. In this paper we present a formulation in terms of {\it dressed} particles, which gives exact Markovian equations. We formulate dressed particles for Poincar\'e nonintegrable systems,  through  an invertible transformation operator $\Lam$ introduced by Prigogine and collaborators. $\Lam$ is obtained by an extension of the canonical (unitary) transformation operator $U$ that eliminates interactions for integrable systems. Our extension is based on the removal of divergences due to Poincar\'e resonances, which breaks time-symmetry.  The unitarity of $U$ is extended to ``star-unitarity'' for $\Lam$. We show that $\Lam$-transformed variables have the same time evolution as stochastic variables obeying  Langevin equations, and that $\Lam$-transformed distribution functions satisfy exact Fokker-Planck equations. The effects of Gaussian white noise  are obtained by the non-distributive property of $\Lam$ with respect to products of dynamical variables. Therefore our method leads to a direct link between dynamics of Poincar\'e nonintegrable systems, probability and stochasticity.
\end{abstract}
\maketitle

%%%%%%%%%%%%%%%%%%%%
% Body  %
%%%%%%%%%%%%%%%%%%%%

%%%%%%%%%%
%SECTION 1
%%%%%%%%%%
\section{Introduction}
\label{sec:Int}
 In classical physics the basic laws are time reversible.
 If we know the Hamiltonian, then we get Hamilton's equations of
 motion which describe the time evolution of the system in a time
 reversible, deterministic way. On the other hand, we see time
 irreversibility and stochastic behavior  everywhere. How to bridge
 the gap between theory and reality has been the subject of many discussions.

The main problem is how to extract  irreversibility and
stochasticity out of Hamilton's equations of motion.
 This will be the subject of this paper. Our approach is an
 extension of canonical transformations to define dressed particles or  quasiparticles \cite{Matt}.

We consider  Hamiltonians that can be written as
\begin{eqnarray}
H = H_0 + \lambda V. \EQN{i-3}
\end{eqnarray}
The first therm $H_0$ describes a set of noninteracting
``bare'' units  while the second $\lam V$ describes their interactions ($\lam$ is a dimensionless coupling constant).
Specifically, we will consider  the one-dimensional Friedrichs model \cite{Friedrichs}, describing a classical harmonic oscillator  (bare particle) coupled to an infinite  set of  bare field modes (heat bath).   This model is closely related to the Caldeira-Leggett model \cite{Caldeira}, which has been extensively  used to study quantum Brownian motion \cite{Gardiner,Fick,Weiss,Nieu,Hu,Calzetta}.

In general, bare particles follow a complicated motion, due to their  interactions. In  order gain a physical insight into their behavior, and also to  simplify the equations of motion, one can introduce a change of phase-space variables (a canonical transformation). The new variables describe renormalized entities, or quasiparticles. After solving the equations for quasiparticles, one may apply the inverse canonical transformation to get the solutions of the original equations of motion.
For the Friedrichs model the quasiparticle consists of the original particle surrounded by a ``dressing'' cloud of field modes.

For integrable systems, one can construct transformations $U$ that completely eliminate the interactions. They bring us to a description in terms of free quasiparticles.  This is readily seen starting with the Liouville equation
\beqa
i\frac{\partial}{\partial t}\rho =  L_H \rho,
\EQN{iLiou}
\eeqa
where $L_H \equiv i\{H,\, \,\}$ is the Poisson bracket with the Hamiltonian.
Similar to \Ep{i-3} the Liouvillian is written as a free term plus interaction,
$L_H = L_0 + \lam L_V$. Applying $U$ on both sides of the Liouville equation we get
\beqa
 i\frac{\partial}{\partial t}U\rho  &=& UL_H U^{-1}U\rho\nonumber\\
\Rightarrow
  i\frac{\partial}{\partial t}\rhob &=& \bar{L_0} \rhob \EQN{iii}
\eeqa
where
\beqa
& &\rhob= U\rho,\quad \bar{L_0}= U L_H U^{-1}.
\eeqa
The transformation $U$ is constructed in such a way that $\bar{L_0}$ has
the same  form as the non-interacting Liouvillian, with renormalized frequencies. \Ep{iii} gives the time evolution of the free dressed particles.

  If $U$ can be constructed through a perturbation expansion in $\lam$, we say the system is integrable in the sense of Poincar\'e. For these systems we can keep a one-to-one correspondence between the original variables and the transformed variables.  With a suitably defined inner product between dynamical variables and ensembles, we can define the hermitian conjugate transformation $U^\dagger$. One then finds that $U$ is unitary: $U^\dagger = U^{-1}$. The transformation $U$ thus preserves the time-reversibility of the original Liouville equation.

Now, if all systems were integrable in Poincar\'e's sense, this would mean that all the phenomena we observe in nature are equivalent to free motion. This would be hard to reconcile with the existence of dissipative phenomena, which are essential for the appearance of  bifurcations and self-organization \cite{FBTB}. However for most systems one cannot construct $U$ by perturbation expansions, due to the appearance of resonances.  Resonances  give vanishing denominators leading to divergences. These divergences were discovered by Poincar\'e, so we will refer to them as Poincar\'e divergences [hereafter, whenever we speak of integrability or nonintegrability, it will be meant in Poincar\'e's sense].

It is precisely for Poincar\'e's nonintegrable systems that we see
irreversible and stochastic behavior, such as  Brownian motion.
One of the main developments of the Brussels-Austin groups led by
I. Prigogine has been to show that one can systematically
eliminate the Poincar\'e divergences by regularization of
denominators \cite{Rosen,PP91,PP96,PP97,OPP}. As a result of this
regularization time-symmetry is broken and one obtains a new type
of transformation $\Lam$ that replaces $U$. This gives a quasiparticle description leading to stochastic or
kinetic equations, such as the  classical Langevin or Fokker-Planck
equations, respectively. To see this, we  operate $\Lam$ on the
Liouville equation
\beqa
i\frac{\partial}{\partial t}\Lam\rho &=& \Lam L_H \Lam^{-1} \Lam\rho\nonumber\\
 \Rightarrow i\frac{\partial}{\partial t}\rhot &=& \tht \rhot
 \EQN{rhot-eq}
\eeqa
where
\beqa
\rhot=  \Lam\rho,\quad \tht= \Lam L_H \Lam^{-1}.
\EQN{rtdef}
\eeqa
$\tht$ is now a collision operator as used in kinetic theory. The $\Lam$ transformation gives a probabilistic description, which is irreducible to trajectories in classical mechanics or wave functions in quantum mechanics. If we integrate out the field variables, \Ep{rhot-eq} becomes, e.g.,  an exact Fokker-Planck operator.  Through the $\Lam$ transformation we can also describe dressed unstable states in quantum mechanics \cite{OPP,POP2001}.   One can define as well an ${\cal H}$-function that has strict monotonic behavior \cite{Rosen}.

For the Friedrichs model we have both integrable and nonintegrable cases, depending on whether the spectrum of the field modes is discrete (finite volume $L$ with periodic boundaries) or continuous ($L\to\infty$). In the first case we have cyclic (although complicated) motion of the particle, as the field comes back to the particle through the periodic boundaries. In the second case, the field does not come back. A Poincar\'e resonance emerges, since  the energy of the particle is embedded inside the continuous spectrum.  The emission of the field from the particle leads to radiation damping.  Conversely, the particle is excited when it absorbs the field.

To understand the breaking of time symmetry, we note that when there are Poincar\'e resonances, i.e. in the limit $L\to\infty$,  the solutions of Hamilton's equations for the bare particle  contain a dominant decaying (Markovian) component oriented either towards the future or the past, or both, depending on the initial conditions.  Taking the well-known Markovian approximation, one finds that the equation of motion for the bare particle is split into two branches, one for $t>0$ and another for $t<0$, corresponding to two semigroups.  As a whole the time reversal invariance of the motion is kept, but if we pick either branch,  time symmetry is broken.  

To obtain this  splitting into two semigroups  for the bare particle we have to make approximations. In contrast,  in terms of the dressed particle defined through $\Lam$, this is an exact property. The analytic continuation of $U$ can be made to either the upper or lower complex frequency planes, giving exact Markovian equations that generate  the $t<0$ or $t>0$ semigroups, respectively. Once we fix the analytic continuation, time symmetry is broken.  

The $\Lam$-transformed functions involve generalized functions, or distributions (examples are the ``Gamow modes'' presented in Sec. \ref{sec:Gam}).  If the initial unperturbed functions formed a Hilbert space, the transformed functions are no more in this Hilbert space.   In  its transformed domain $L_H$ behaves as the dissipative collision operator $\tht$ with complex eigenvalues \cite{PP96,PP97}.

In contrast to $U$, $\Lam$ is no more unitary. Instead, it is ``star unitary'' \cite{Rosen,OPP}. Furthermore, while $U$ is distributive with respect to multilplication, $\Lam$ is non-distributive. As we will see, these properties allow us to describe damping and fluctuations associated with noise.

A basic requirement on  $\Lam$ is that it is invertible. This is connected with the star unitarity of this transformation [see comments below \Ep{StarUnit}]. In addition to this,
our construction of $\Lam$ is based on the following requirements:

\smallskip
\noindent (1) The $\Lam$ transformation is obtained by  analytic
continuation of the unitary transformation $U$. When there are no resonances, $\Lam$  reduces to $U$. \newline\
(2) $\Lam$ preserves the measure of the phase space.\newline
(3) $\Lam$ maps real variables to real variables.\newline
(4) $\Lam$ is analytic with respect to the coupling constant $\lam$ at $\lambda=0$. \newline
(5) $\Lam$ leads to closed Markovian kinetic equations.
\smallskip

We  will focus on the  dynamical observables of the particle. The action of $\Lam$ will be restricted to the subset of phase space functions depending only on the particle degrees of freedom. Within this subset we obtain an
exact and invertible  $\Lam$ transformation.

Furthermore, we will consider the thermodynamic limit of the field modes, where the total energy of the field  is an extensive variable. Then the average action $\bra J_k\ket$ of each field mode $k$ satisfies \cite{Valeri}
\beqa
\bra J_k\ket \sim O(L^0)
\EQN{thlim}
\eeqa
for $L\to\infty$. The  total energy of the field
\beqa
E_{\rm f} =\sum_k \ome_k \bra J_k\ket \to \frac{L}{2\pi} \int dk\, \ome_k \bra J_k\ket \sim O(L)
\EQN{thlim'}
\eeqa
is proportional to the volume $L$. [This does not necessarily imply that the field is Gibbsian]. The existence of the thermodynamic limit requires an initially random distribution of the phases of the  field modes \cite{PP2000}.

A different situation occurs if the total energy of the  field  is in a non-extensive variable.  Then we have $\bra J_k\ket \sim O(L^{-1})$, i.e., we have a vanishing energy density. We will not consider this case in this paper.

In the extensive case, in addition to the damped oscillation, the particle undergoes an erratic motion due to the excitation caused by the  field.   This erratic motion includes a Brownian motion component, which is Markovian. The initial randomness of the phases of the field modes is a  necessary condition for the appearance of Brownian motion. In addition it is essential that the field {\it resonates} with the particle. We need Poincar\'e resonances. Under these conditions $\Lam$ permits us to isolate the damping and the Brownian component of the motion.

 Our approach can also
be formulated in terms of complete sets of projection operators $\Pi^{(\nu)}$, that permit to decompose dynamics into a set of orthogonal ``subdynamics'' (see Sec. \ref{sec:Lambda}). Essentially, we introduce a generalized basis that permits us to analyze the motion in  terms of strictly Markovian components. In our case, we study the component that describes Brownian motion. The other components, clumped together, give what is usually called non-Markovian (memory) effects \cite{PP97}. The Brownian component is independent of the initial correlations between the particle and the bath, and in this sense, it has a ``universal'' character.

The results presented here are based on Refs. \cite{OPP,POP2001}, where we constructed $\Lam$ for the quantum Friedrichs model.
The main subject in these papers was the decay of unstable particle states. We  showed that the $\Lam$ transformation permits us to isolate the exponential (Markovian) component of the decay, which occurs when the energy of the field is non-extensive. The remaining (non-Markovian) component gives the Zeno effect  \cite{SudZ} and long tails \cite{Kalf}, which are conneted to the appearance of a dressing cloud around the bare particle. The dressed unstable state defined through $\Lam$  has a real average energy and gives an uncertainty relation between the lifetime and energy (see also \cite{PP88}).  Similar considerations can be applied in classical mechanics \cite{rdamping}.

The present paper is organized as follows. In Secs.  \ref{sec:model} -  \ref{sec:bare}
 we introduce the Friedrichs model and we discuss the equations of motion of the bare particle. In the subsequent Sections   we study the evolution of renormalized (dressed) particle variables.
 We consider first (Sec. \ref{discrete}) the integrable case where the spectrum of the field is discrete. We can then define the renormalized variables through  the unitary transformation $U$.   In  the continuous spectrum limit the system becomes nonintegrable. In Sec. \ref{sec:Gam}, as a first step to introduce $\Lam$,  we extend the renormalized particle modes in the discrete case to the decaying ``Gamow'' modes in the nonintegrable case. In Sec. \ref{sec:Lambda} we construct $\Lam$. In Sec. \ref{sec:lang} we show the correspondence between the solution of Langevin equation (with Gaussian white noise) and the $\Lam$ transformed variables. Finally, in Sec. \ref{sec:fp-drv} we derive a Fokker-Planck equation for the $\Lam$ transformed distribution function. Details of calculations are given in the Appendices.

%%%%%%%%%%%%%%%%%%%%%%
% Section 2
%%%%%%%%%%%%%%%%%%%%%%

\section{The classical Friedrichs model}
\label{sec:model}
We consider a classical system consisting of a harmonic oscillator coupled to a classical scalar field in one-dimensional space.
A quantum version of this model has been studied by Friedrichs \cite{Friedrichs}, among others.

We write the Hamiltonian of the system in terms of the bare oscillator and field modes $\qb_1$ and $\qb_k$,
\begin{eqnarray}
H
  = \omega_1 \qb^*_1 \qb_1 + \sum_k \omega_k \qb^*_k \qb_k + \lambda \sum_k
\Vb_k (\qb_1^* \qb_k + \qb_1 \qb_k^* ), \nonumber\\
\EQN{ham}
\end{eqnarray}
with a given constant frequency $\ome_1 >0$ for the harmonic oscillator (particle), $c=1$ for the speed of light, and  $\ome_k = |k|$ for the field. When $\lam$ is small we can treat the interaction potential as a perturbation. We assume the system is in a one-dimensional box of size $L$ with  periodic boundary conditions. Then the spectrum of the field is discrete, i.e., $k= 2\pi j/L$ where $j$ is an integer. We assume that
\beqa
\ome_1\ne\ome_k,  \quad {\rm for\, all\,}  k
\EQN{ASS}
\eeqa

The volume dependence of the interaction $V_k$ is given by
\beqa
\Vb_k = \sqrt{\frac{2\pi}{L}} \vb_k
\eeqa
where $\vb_k = O(1)$. We assume that $\vb_k$ is real and even: $\vb_k=\vb_{-k}$. Furthermore, we assume that for small $k$
\beqa
\vb_k \sim \ome_k^{1/2}. \EQN{3}
\eeqa
An example is the Drude-Ullersma form \cite{Nieu}:
\beqa
\vb_k = \frac{\ome_k^{1/2}}{1 + \ome_k^2/\ome_M^2}, \EQN{Drude}
\eeqa
where $\ome_M$ is the cutoff frequency of the bath.

 To deal with the continuous spectrum of the field we take the limit
$L \rightarrow \infty$. In this limit we have
\begin{eqnarray}
\frac{2\pi}{L} \sum_k \rightarrow \int dk,\;\;\;
\frac{L}{2\pi}\delta_{k,0} \rightarrow \delta(k).
\end{eqnarray}
We will often use the summation sign with the understanding that we replace it by an integral in the limit $L\to\infty$.

 The bare modes  $\qb_1$, $\qb_k$ satisfy the Poisson bracket relation
\begin{eqnarray}
 i\{\qb_{\alpha},\; \qb^*_{\beta} \} = \delta_{\alpha\beta}.
 \EQN{Poissonq}
\end{eqnarray}
where
\begin{equation}
   i\{f,g\} = \sum_{r} \Big[
      \frac{\partial f}{\partial \qb_r}  \frac{\partial g}{\partial \qb_r^*}
   -  \frac{\partial g}{\partial \qb_r}  \frac{\partial f}{\partial \qb_r^*}
   \Big]
    \EQN{PBde}
\end{equation}
[the sum includes the discrete index $r=1$ as well as the running index $r=k$].
The bare modes are related to the position $x_1$ and the momentum $p_1$ of the particle as
\begin{eqnarray}
\qb_1 &=& \sqrt{\frac{m \omega_1}{2}} (x_1 + \frac{i p_1}{m\omega_1}), \\
x_1 &=& \frac{1}{\sqrt{2m\ome_1}}(\qb_1+\qb^*_1), \nonumber\\
 p_1 &=& -i\sqrt{\frac{m\ome_1}{2}}(\qb_1-\qb^*_1) \EQN{x1p1}
\end{eqnarray}
and to the field $\phi(x)$ and its conjugate field $\pi(x)$ as
\begin{eqnarray}
\phi(x)&=& \sum_{k} \big(\frac{1}{2\ome_k L}\big)^{1/2}(\qb_k e^{ikx}+\qb^*_k e^{-ikx}), \\
\pi(x)&=& -i \sum_k \big(\frac{\ome_k}{2L}\big)^{1/2} (\qb_k e^{ikx}-\qb^*_k e^{-ikx}).
 \EQN{phixpix}
\eeqa
The field $\phi(x)$ corresponds to the transverse vector potential in electromagnetism, while $\pi(x)$ corresponds to the transverse displacement field. Our Hamiltonian can be seen as  a simplified version of a classical dipole molecule interacting with a classical radiation field in the dipole approximation \cite{Passante}. For simplicity we neglect  the interactions proportional to $\qb_1\qb_k$ and $\qb^*_1 \qb^*_k$, which correspond to ``virtual processes'' in quantum mechanics. This approximation corresponds to the so-called rotating wave approximation \cite{cohen, Gardiner}.  If we incorporate the virtual processes, then we obtain the classical version of  the Caldeira-Leggett model.

 We note that we have an $\ome_k = \ome_{-k}$ degeneracy in our Hamiltonian. To avoid some complexity due to this degeneracy, we rewrite our Hamiltonian in terms of new bare modes as \cite{rdamping}
\beqa
H = \ome_1 q_1^*q_1 + \sum_k \ome_k q^*_k q_k + \lam \sum_k V_k(q^*_1 q_k + q_1q^*_k),
\eeqa
where
\beqa
q_1 \equiv \qb_1, \quad q_k \equiv \left\{ \begin{array}{ll}
                               (\qb_k + \qb_{-k})/\sqrt{2}, \,\, & \mbox{for $k > 0$,} \\  (\qb_k-\qb_{-k})/\sqrt{2}, \,\, & \mbox{for $k \le 0$,}
\end{array} \right.
\eeqa
\beqa
&&V_k \equiv \left\{ \begin{array}{ll}
                    \sqrt{2}\Vb_k, \,\, & \mbox{for $k>0$} \\
                    0,    \,\, & \mbox{for $k \le 0$,} \end{array}\right. \\
&&v_k = \sqrt{\frac{L}{2\pi}} V_k.
\eeqa
 In this form the mode $q_k$ with  negative $k$ argument  is completely decoupled from the other degrees of freedom.
The new bare modes also satisfy equations (\ref{Poissonq}), (\ref{PBde}).

In the subsequent sections we will use the following notations.
We define action and angle variables $J_s,\alp_s$ through the relation
\begin{eqnarray}
q_s = \sqrt{J_s} e^{-i\alp_s}, \quad s=1,\,k
\EQN{aang}
\end{eqnarray}

We define $\Gam$ as the set of all modes,
 \beqa
 \Gam \equiv (q_1, q_1^*,...,q_k,q_k^*,...)
 \eeqa
and $\Gam_s = (q_s, q_s^*)$, with $s=1,k$, as the set of particle or field modes. We will also denote $\Gam_f$
as the set of all field modes $\Gam_f = \{\Gam_k\}$. We use the notation $d\Gamma$ for the phase space volume element
and $d\Gam_1$, $d\Gam_f$ for the particle and field components of $d\Gam$, respectively
\beqa
&&d\Gam = d\Gam_1 d\Gam_f \nonumber\\
& &d\Gam_1 = dJ_1{d\alp_1},\quad d\Gam_f =
\prod_k dJ_k {d\alp_k}. \eeqa 
We define as well
 \beqa
 & &\delta (\Gam-\Gam') \nonumber \\
 & &\equiv \delta (J_1 - J_1')\delta (\alp_1-\alp_1')
 \prod_k \delta (J_k - J_k')\delta (\alp_k-\alp_k'). \nonumber \\
 \eeqa
We consider ensemble averages as inner products:
\begin{equation}
    \bra F \ket = \bbra F|\rho \kket = \int d\Gamma F(\Gamma)^* \rho(\Gamma).
      \EQN{ensav}
\end{equation}
For an operator $O$ the Hermitian conjugate is defined by
\begin{equation}
    \bbra F |O\rho\kket = \bbra \rho|O^\dagger F\kket^*.
      \EQN{Hc}
\end{equation}

As mentioned in the Introduction, for our model we can have both integrable ($L$ finite)  and nonintegrable
cases ($L\to\infty$). In the first case there are no
resonances (see \Ep{ASS}) and as we will see, the  the system is
integrable in the sense of Poincar\'e. In the second case the
system can become nonintegrable in Poincar\'e's sense, due to the
emergence of the resonance $\ome_1=\ome_k$ between the frequencies
of the particle and the field. This distinction is essential in
our construction of dressed particle modes. Before coming to this,
we   will briefly consider the equations of motion for the bare
modes.

%%%%%%%%%%%%%%%%%%%%
% Section
%%%%%%%%%%%%%%%%%%%%

\section{Equations of motion of the bare particle modes}
\label{sec:bare}

The dynamical equations of motion of an oscillator coupled to a field have been studied by many authors, mainly using the quantum Caldeira-Leggett model \cite{Caldeira,Gardiner,Fick,Weiss,Nieu}. Here we will write the equations for the Friedrichs model.  In contrast to the phenomenological equations describing Brownian motion \cite{Langevin}, these  equations have memory terms (i.e. they have time-dependent damping and diffusion coefficients),   and the time evolution they generate forms a group, since they are equivalent to Hamiltonian dynamics.

%%%%%%%%%%%%%%%%%%
\subsection{Non-Markovian Langevin equation}
%%%%%%%%%%%%%%%%%%%
Starting from the Hamiltonian equations
\begin{eqnarray}
q_s(t) = \exp(i L_H t) q_s(0) \Rightarrow {\dot q}_s(t) = i L_H
q_s(t), \EQN{stand1}
\end{eqnarray}
we can obtain the exact time evolution of the modes $q_r$ as
\begin{eqnarray}
 q_s(t) &=&  \sum_r f_{sr}(t) q_r(0)
 \EQN{sol1}
\end{eqnarray}
where $f_{rs}(t)$ are complex functions (see Appendix {\ref{apx:frs}).

 We will focus our attention on the particle modes:
\begin{eqnarray}
 q_1(t) &=&  f_{11}(t) q_1(0) +  \sum_k f_{1k}(t) q_k(0),  \nonumber\\
 {\dot q}_1(t) &=&  {\dot f}_{11}(t) q_1(0) +  \sum_k {\dot f}_{1k}(t)
 q_k(0).
 \EQN{sol2}
\end{eqnarray}
Solving for $q_1(0)$ in the first equation and replacing the result in the second equation we get
\begin{eqnarray}
 {\dot q}_1(t) &=&  - i z_1(t) q_1(t) + R(t)
 \EQN{stand4}
\end{eqnarray}
where
\begin{eqnarray}
 z_1(t)&=& i \frac{\pard}{\pard t} \ln f_{11}(t) , \EQN{stand5}\\
 R(t) &=& \sum_k h_k(t) q_k(0),\EQN{stand5'}\\
 h_k(t) &=& {\dot f}_{1k}(t) + i z_1(t) f_{1k}(t). \EQN{stand5''}
\end{eqnarray}
\Ep{stand4}  is a non-Markovian  equation, because of the time dependence of the coefficients.  The function $z_1(t) \equiv \omet_1(t) - i \gamma(t)$ gives the instantaneous frequency $\omet_1(t)$ and damping rate $\gamma(t)$ of the oscillator [we note that  damping appears only in the non-integrable case]. $R(t)$ is an erratic function, since it depends on the initial states of all the field modes $q_k(0)$ (see Appendix \ref{apx:thnoise}). It plays the role of noise.  In general, this is colored noise, as the function $R(t)$ has  memory in the  auto-correlation,
\begin{eqnarray}
\bra R^*(t) R(t') \ket \ne 0 \quad {\rm for}\,\, t\ne t',
\EQN{stand6}
\end{eqnarray}
where $\bra\, \ket$ means ensemble average.

%%%%%%%%%%%%%%%%%%%
\subsection{Non-Markovian Fokker-Planck equation}
%%%%%%%%%%%%%%%%%%%
We can also derive a non-Markovian equation for the particle distribution function
 \beqa
 \rho_1 (\Gam_1, t) \equiv \int d\Gam_f\, \rho (\Gam,t)
 \eeqa
which allows us to calculate  averages of functions $G(\Gam_1)$ depending only on the particle modes.
We assume that $G(\Gam_1)$ is a smooth real function of $\Gam_1$ that vanishes  at  $|q_1|=\infty$ and is expandable in the infinite series
\beqa
 G(q_1, q_1^{*}) = \sum_{m=0}^\infty \sum_{n=0}^\infty G_{mn} q_1^{*m}
 q_1^n.
 \EQN{mn7}
\eeqa
We assume as well  that $\rho(\Gam,t)$ is factorized at $t=0$ into independent particle and
field mode functions and that the field distributions depend only on the actions. In other words, we have
\beqa
\rho(\Gam,0) = \rho_1(\Gam_1,0) \prod_{k}\rho_k(J_k). \EQN{nfok2}
\eeqa
Our final assumption is that the volume of the system is large, so we neglect terms of order $1/L$. This approximation becomes exact in the continuous spectrum limit $L\to\infty$, i.e., in the nonintegrable case. We consider the extensive case
discussed in the Introduction.

Based on \Ep{stand4} we then obtain the non-Markovian  equation (see Appendix \ref{apx:nmfok})
\beqa
 \frac{\partial}{\partial t} \rho_1(\Gam_1,t) &=&   \left\{ iz_1(t) \frac{\partial}{\partial q_1} q_1 -iz_1^*(t) \frac{\partial}{\partial q^*_1} q^*_1
\right.
\EQN{nfok4} \\
 &+& \left. D(t) \frac{\partial^2}{\partial q_1 \partial q^*_1} \right\}\  \rho_1(\Gam_1,t),
\nonumber
\eeqa
where
\beqa D(t) = \sum_k \left[ -iz_1^*(t) + iz_1(t) +
\frac{\pard}{\pard t} \right] |f_{1k}(t)|^2 \bra J_k \ket.
\EQN{Dtres2} \eeqa %%
\Ep{nfok4}  is of the Fokker-Planck type, but with time dependent coefficients.

The equations derived in this section are  reminiscent of the
phenomenological equations for Brownian motion. However, the
phenomenological equations have quite important differences: they are
Markovian, they  break time-symmetry and they describe stochastic
processes.  One can derive the phenomenological equations using
approximations, such as the Markovian approximation. This is shown
in Appendix \ref{apx:mark}.

 In the rest of the paper we will study the dynamical evolution of renormalized modes.  In contrast to the bare modes,  the renormalized modes obey exact   equations having the same evolution as the phenomenological equations. As a preparation, we first consider the integrable case.

%%%%%%%%%%%%%%%%%%%%%%
% Section 3
%%%%%%%%%%%%%%%%%%%%%%

\section{Unitary transformation for integrable case}
\label{discrete}

In this Section, we present the properties of the canonical transformation $U$ that diagonalizes the Hamiltonian in the discrete spectrum case, when the size of the  box $L$ is finite.  Later we will extend $U$ to $\Lam$ through analytic continuation, for $L\to\infty$.
In the integrable case we can find  renormalized  modes  $\Qb_{s}$, $\Qb^*_{s}$  that   diagonalize the Hamiltonian through  $U$.
The new  modes  are related to the bare modes as
\begin{eqnarray}
\Qb_{s} =U^{\dagger}q_{s} \quad {\rm for}\,\,  s = 1, k.
 \EQN{Qb-q}
\end{eqnarray}
in  one-to-one correspondence.  The operator $U$ is unitary $U^{-1} = U^{\dagger}$.

The Hamiltonian is diagonalized as
\begin{eqnarray}
 H = \sum_s \omeb_s  \Qb^*_s \Qb_s
 \EQN{Hdiag}
\end{eqnarray}
where $\omeb_\alp$ are renormalized frequencies.

The new  modes satisfy the Poisson bracket relation
\begin{eqnarray}
 i\{\Qb_{r},\; \Qb^*_{s} \} = \delta_{rs}.
 \EQN{PoissonQ}
\end{eqnarray}
Since the interaction is bilinear in the bare modes, the new  modes can be found explicitly through a linear superposition of the bare modes \cite{rdamping}. For the particle we obtain, from the equation $i\{H,\Qb_1\} = - \omeb_1 \Qb_1$,

\begin{eqnarray}
\Qb_1 = \Nb^{1/2}_1 (q_1 + \lam \sum_k \cb_k q_k) \EQN{Qb_1-6}
\eeqa
where
\beqa
& &\cb_k \equiv \frac{V_k}{\omeb_1-\ome_k}, \\
& &\Nb_1 \equiv (1+\xib)^{-1},\quad \xib \equiv \lam^2 \sum_k \cb_k^2.
 \EQN{cbN1}
\eeqa
The renormalized frequency $\omeb_1$ is given by the root of the equation
\begin{eqnarray}
 \eta (\omeb_1)= 0, \quad
\eta (z) &\equiv& z - \ome_1 - \sum_{k'} \frac{\lambda^2
|V_{k'}|^2 }{z -\ome_{k'}}
\EQN{eta-root}
\end{eqnarray}
that reduces to $\ome_1$ when $\lam=0$.
For the field modes one can also find explicit forms (see Appendix \ref{apx:frs}).

The perturbation expansion of \Ep{Qb_1-6} yields
\begin{eqnarray}
\Qb_1 = U^{\dagger} q_1 = q_1 +  \sum_k {\lam V_k \over \ome_1 - \ome_k} q_k + O(\lam^2)
\EQN{pertQb1}
\eeqa

When the spectrum is discrete, the denominator never vanishes; each term
in the perturbation series is finite.  This implies integrability in the sense of Poincar\'e: $U$ can be  constructed by a perturbation series in powers of $\lam^n$ with $n\ge 0$ integer. In other words, $U$ is analytic at $\lam=0$.

Since the transformation $U$ is canonical, it  is distributive with respect to multiplication
\beqa
U^{\dagger}q_{r}q^*_{s} = [U^{\dagger}q_{r}][U^{\dagger}q^*_{s}]= \Qb_{r}\Qb^*_{s}
\EQN{distrib}
\eeqa
Hence we have
\begin{eqnarray}
 UH = U[ \sum_s \omeb_s  \Qb^*_s \Qb_s  ]
 =  \sum_s \omeb_s q^*_s q_s  = {\bar H_0} \EQN{Hob}
\end{eqnarray}
The transformed Hamiltonian $UH$ has the same form of the unperturbed Hamiltonian $H_0$, with renormalized frequencies.

The canonical transformation can also be introduced on the level of statistical ensembles $\rho$, as shown in the Introduction.  In \Ep{iii}  we have
\beqa
& & \bar{L}_0 \rhob = iU\{ H, \rho\} = i\{UH, U\rho\} \nonumber \\
& &= \Bigg[ \sum_s \omeb_s (q^*_s\frac{\partial }{\partial q^*_s} - q_s\frac{\partial}{\partial q_s})\Bigg]U\rho
 \EQN{LUp}
\eeqa
where in the second equality we used \Ep{Hob}  and the property of preservation of the Poisson bracket by canonical transformations \cite{Balescu}. Hence the transformed Liouvillian $\bar{L}_0$ does not contain any interaction terms. Ensemble averages over this transformed density function $\rhob$ can thus be easily calculated. For example for
\beqa
 i \frac{\partial}{\partial t} \bbra  x_1| U|\rho\kket = i \frac{\partial}{\partial t}\bra {\bar x}_1 \ket
\EQN{x1int}
\eeqa
and similarly for $\bra {\bar p}_1\ket$ we get, after
substituting \Ep{x1p1} and integrating by parts,
\beqa
\frac{\partial}{\partial t}\bra {\bar x}_1 \ket = \frac{1}{{\bar m}}\bra {\bar p}_1 \ket ,\quad \frac{\partial}{\partial t} \bra {\bar p}_1 \ket = -{\bar m}\omeb_1^2 \bra {\bar x}_1 \ket.
\EQN{x1dot}
\eeqa
These are the equations for the free harmonic oscillator (with renormalized frequency $\omeb_1$ and renormalized mass ${\bar m} = m \ome_1/\omeb_1$).
The interaction with the field is eliminated.

Note that the normal modes are eigenfunctions of the Liouvillian ${\bar L_0}$,
\beqa
 {\bar L_0} q_1 =  -\omeb_1q_1, \qquad {\bar L_0} q_1^* = \omeb_1 q_1^*.
 \EQN{qeig}
\eeqa
This leads to
\beqa
 L_H \Qb_1 =  -\omeb_1 \Qb_1, \qquad L_H \Qb_1^* = \omeb_1 \Qb_1^*.
 \EQN{Qbeig}
\eeqa
For products of  modes we have
\beqa
{\bar L_0} q_1^{*m} q_1^n
  &=&  [(m - n)\omeb_1]\, q_1^{*m} q_1^n, \nonumber \\
 L_H \Qb_1^{*m} \Qb_1^n
  &=&  [(m - n)\omeb_1]\, \Qb_1^{*m} \Qb_1^n .
 \EQN{qeig2}
\eeqa

Finally, we note that from distributive property \Ep{distrib} we have
\beqa
U^{\dagger} q_1^{*m} q_1^n = (U^{\dagger} q_1^{*m}) (U^{\dagger}q_1^n).
 \EQN{Uqmn}
\eeqa
%%

%%%%%%%%%%%%%%%%%%%%%%
% Section
%%%%%%%%%%%%%%%%%%%%%%
\section{Nonintegrable case: Gamow modes}
\label{sec:Gam}

Now we consider the continuous spectrum case, where   the particle frequency $\ome_1$ is inside the range of the continuous spectrum $\ome_k$. In this case, by analytic continuation of $\Qb_1$ and $\Qb_1^*$ we can get new modes which are eigenfunctions of the Liouvillian with complex eigenvalues. These modes are called Gamow modes. Gamow states  have been previously introduced  in quantum mechanics to study unstable states  \cite{Sudarshan}-\cite{AGKPP}. In classical mechanics, Gamow modes have been introduced in Ref. \cite{rdamping}.  In this Section we present the main properties of Gamow modes, which will be used for the construction of $\Lam$.

When we go to the continuous limit
we restrict the strength of the  coupling constant $\lambda$ so that
\begin{eqnarray}
\int\ dk \frac{\lambda^2 |v_k|^2 }{ \ome_k} < \ome_1,
 \EQN{weakl}
\end{eqnarray}
Then the harmonic oscillator becomes unstable. In this case we have radiation damping.  If \Ep{weakl} is not satisfied, then we go outside the range of applicability of the ``rotating wave approximation'' (see comment after \Ep{phixpix}) as the Hamiltonian becomes not bounded from below, and gives no radiation damping \cite{ppri}.

 In the continuous spectrum case, divergences appear in the construction of $U$, due to resonances. For example, the denominator  in \Ep{pertQb1}
may now vanish at the Poincar\'e resonance $\ome_1 = \ome_k$. We have a divergence in the perturbation expansion in $\lam$. To deal with this divergence, we regularize the denominator by adding an infinitesimal $\pm i\eps$. Then we get
\begin{eqnarray}
Q_1 =  q_1 + \sum_k {\lam V_k \over \ome_1 - \ome_k \pm i\eps} q_k + O(\lam^2).
\EQN{pertQ1}
\eeqa
In the continuous limit the summation goes to an integral.  We take the limit $L\to \infty$ first and $\eps \to\infty$ later. Then the denominator can be interpreted as a distribution under the integration over $k$
\begin{eqnarray}
{1\over \ome_1 - \ome_k \pm i\eps} \to {\cal P} {1\over \ome_1 - \ome_k} \mp  i\pi\del(\ome_1-\ome_k)
\EQN{Pdel}
\eeqa
where $\cal P$ means principal part.

The introduction of $i\eps$ in the continuous limit is related to a change of the physical situation. In the discrete case the boundaries of the system cause periodicity in the motion of the particle and the field. In contrast, in the continuous case the boundaries play no role.
In the continuous limit we can have  damping of the particle, as the field emitted from the particle
goes away and never comes back. And we can have Brownian motion, due to the interaction with
the continuous set of field modes. The continuous limit may be well approximated by  a discrete system during time scales much shorter than the time scale for which the field goes across the boundaries.

In the continuous limit we can have damping of the particle either
toward the future or toward the past. This corresponds to the
existence of the two branches  $\pm i\eps$ in \Ep{pertQ1}.
Breaking of time symmetry is connected to resonances \cite{PP88a}.

As shown in \cite{PPT}, continuing the perturbation expansion (\ref{pertQ1}) to all orders one obtains new
renormalized modes (Gamow modes) associated with the complex frequency
\beqa
 z_1 \equiv \omet_1 -i\gamma
 \EQN{z1def}
\eeqa
or its complex conjugate $z_1^*$. Here $\omet_1$ is the renormalized frequency of the particle, and $2\gamma > 0$ is the damping rate.
The complex frequencies are  solutions of the equation
\beqa
 \eta^\pm(\ome) = \ome -\ome_1 - \int dk \frac{\lambda^2 v_k^2}{(z-\ome_k)^\pm_{\ome}}
 =0.
 \EQN{etaz1}
\eeqa
The $+$ ($-$) superscript
indicates that the propagator is first evaluated on the upper (lower)
half plane of $z$ and then analytically continued to $z=\ome$.

The new modes for the $-i\eps$ branch in \Ep{pertQ1} are given by
\begin{eqnarray}
\Qt_1 &=& N_1^{1/2} [ q_1 +\lambda \sum_k c_k q_k], \\
 c_k &=&\frac{ V_k}{(z-\ome_k)^+_{z_1}},\quad N_1 = (1+ \lambda^2 \sum_k c_k^2)^{-1}  \EQN{Q*1}
\end{eqnarray}
and its complex conjugate, satisfying
\begin{equation}
L_H \Qt_1 =  -z_1 \Qt_1, \qquad L_H\Qt_1^* = z_1^* \Qt_1^*.
\EQN{LHAt}
\end{equation}
The mode $\Qt_1^*$ decays for $t>0$ as
\begin{equation}
e^{iL_H t} \Qt_1^* = e^{i z_1^* t} \Qt_1^* = e^{(i\omet_1 -\gamma)t} \Qt_1^*
\EQN{At1dec}
\end{equation}
(and similarly $\Qt_1$).

The modes for the $+i\eps$ branch are given by
\begin{eqnarray}
Q_1^*  =  N_1^{1/2} [q_1^* +\lambda \sum_k c_k q_k^*]
 \EQN{A*1}
\end{eqnarray}
and its complex conjugate, satisfying
\begin{equation}
L_H  Q_1^* =  z_1 Q_1^*, \qquad L_H Q_1 = -z_1^*Q_1. \EQN{LHA}
\end{equation}
These modes decay for $t<0$.

The modes we have introduced have quite different properties from the usual canonical variables.
Their Poisson brackets vanish
\beqa i\{Q_1, Q_1^*\} =  i\{\Qt_1, \Qt_1^*\}  = 0.
 \EQN{PBA}
\eeqa
However the modes  $\Qt_1$ and $Q_1^*$ are duals; they form a generalized canonical pair
\beqa i\{\Qt_1, Q_1^*\} =1.
 \EQN{At1A}
\eeqa
This algebra corresponds to an extension of the usual Lie algebra including
dissipation. An analogue of this algebra has been previously studied
in quantum mechanics \cite{Sudarshan}-\cite{AGKPP}, in terms of non-Hilbertian  bras and kets.

%%%%%%%%%%%%%%%%%%%%%%
% Section
%%%%%%%%%%%%%%%%%%%%%%
\section{The  $\Lam$ transformation}
\label{sec:Lambda}

Using the above results we now introduce $\Lam$. In this paper we will restrict the action  of $\Lam$ to products of particle modes of the form $q_1^{*m} q_1^n$.   This will be enough  to calculate renormalized functions of the particle variables (expandable in monomials), which will lead us to the Langevin and Fokker-Planck equations.  The action of $\Lam$ on more general functions, including field modes will be considered elsewhere  (see also \cite{OPP,rdamping}).

\subsection{Defining $\Lam$ through its action on particle modes}
%%%%%%%%%%%%%%%%%%%%%%%%%%%%%%%%%%%
First recall that in the integrable case the renormalized particle modes are related to
the original modes as
\beqa
\Qb_1 = U^\dagger q_1, \qquad
\Qb_1^* =  U^\dagger q_1^*
 \EQN{Lam_1}
\eeqa
For products of modes we have as well the relation (\ref{Uqmn}). In the continuous spectrum limit we come to the nonintegrable case.
As seen in the previous Section, we eliminate Poincar\'e divergences in single renormalized particle modes by analytic continuation of frequencies to the complex plane (i.e. $\omeb_1$ goes  to $z_1$) leading to Gamow modes.  There are two branches for the continuation, namely
\beqa
\Qb_1 \Rightarrow  \left\{ \begin{array}{c}
                     \Qt_1  \\
                     Q_1  \\
                      \end{array} \right. . \EQN{Lam_2}
\eeqa
Corresponding to these extensions, we introduce $\Lam$, the extension of $U$ in \Ep{Lam_1},
\beqa
\Qt_1 &=& \Lam^{\dagger} q_1, \quad \Qt_1^* = \Lam^{\dagger} q_1^*, \nonumber\\
Q_1 &=& \Lam^{-1} q_1, \quad Q_1^* = \Lam^{-1} q_1^*.
 \EQN{ABL}
\eeqa
These relations partially define $\Lam$, by its action on single
particle modes (a more complete definition is given below).   This
definition satisfies the requirements (1), (3) and (4) given in
the Introduction. We will comment on the remaining requirements
(2) and (5) below.
  Note that $\Lam^\dagger\ne \Lam^{-1}$ is not unitary. Instead, it is ``star-unitary,''
\beqa \Lam^{-1} = \Lam^{\star}.
 \EQN{StarUnit}
\eeqa
In our case, where we  restrict the action of $\Lam$ to particle modes, star conjugation has a simple meaning. It simply
means taking hermitian conjugation and changing $i\eps\Rightarrow -i\eps$, so we have, e.g.,  $[\Lam^{\star} (i\eps)]q_1 = [\Lam^\dagger (-i\eps)] q_1$. For the general definition of star conjugation, see \cite{PPT,OPP} and references therein.

Due to star-unitarity, the existence of the star-conjugate transformation $\Lam^{\star}$ guarantees the existence of the inverse $\Lam^{-1}$.

As mentioned above we are interested not only in the renormalized modes, but also the renormalized products of modes,
\beqa
 \Lam^{\dagger} q_1^{*m} q_1^n, \quad \Lam^{-1} q_1^{*m} q_1^n.
 \EQN{Lqmn}
\eeqa

For the  integrable case, renormalized products of modes can be easily calculated
thanks to the distributive property (\ref{Uqmn}). However, as shown below, for the nonintegrable case products of Gamow modes give new Poincar\'e divergences. Hence, due to the requirement (4) stated in the Introduction, the $\Lam$ transformation has to be non-distributive.
This means that we still have to define the action of $\Lam$ on products of particle modes.

Let us first consider the transformed product $\Lamd q_1^*q_1$. Later we will generalize this to obtain the expressions (\ref{Lqmn}). If $\Lamd$ were distributive, $\Lamd q_1^*q_1$ could be expressed as the product $\Qt_1^*\Qt_1 = (\Lam^{\dagger}q^*_1)(\Lam^{\dagger}q_1)$. However, as we show now, this expression gives Poincar\'e divergences in the thermodynamic limit. We have
\beqa
& & \Qt_1^*\Qt_1 \nonumber\\
&=& |N_1| (q_1^* +\lambda \sum_k c_k^* q_k^*)(q_1 +\lambda \sum_k c_k q_k)\nonumber \\
&=& |N_1|(q^*_1q_1 + \lam q^*_1 \sum_k c_k q_k + \lam q_1 \sum_k c^*_k q_k^* \nonumber \\
&+& \lam^2 \sum_{k,k'}' c_k^* c_{k'} q^*_k q_{k'} + \lam^2 \sum_k |c_k|^2 q^*_k q_k). \EQN{Q*1Q1}
\eeqa
where the prime in the summation means $k\ne k'$.  Going to the continuous limit  and taking the ensemble average with an ensemble $\rho$ the last term becomes
\beqa
\sum_k |c_k|^2 \bra q^*_k q_k \ket \to  \int dk   \Big|\frac{\lam  v_k }{(z-\ome_k)^+_{z_1}}\Big|^2 \bra J_k \ket
\EQN{ck2'}
\eeqa
where $\bra J_k\ket = \bbra q^*_k q_k|\rho\kket$. This term has a non-vanishing
finite value in the limit $L\to\infty$ if \Ep{thlim} is satisfied. Furthermore, if the ensemble $\rho$ belongs to the class of ensembles with $\delta$-function singularities in the wave number $k$, then \Ep{ck2'} is non-negligible as compared to  the average of the $q^*_k q_{k'}$ term in \Ep{Q*1Q1}. For this class of ensembles  the point contribution $k=k'$ is as important as the integration over $k'$ \cite{IP62,PP96,PP97}:
\beqa
\sum_{k'} \bra q_k^* q_{k'}\ket  \sim  \bra J_k\ket \sim O(L^0).
\EQN{dsing}
\eeqa
(see Appendix \ref{app:dsing}). This type of ensembles with $\delta$-function singularities is by no means atypical. An example of this class of ensembles is the  Gibbs distribution.  For ensembles in this class, we have well defined intensive and extensive variables in the thermodynamic limit \cite{IP62}.

To lowest order we have in \Ep{ck2'},
\beqa
\frac{\lam  v_k }{(z-\ome_k)^+_{z_1}} = \frac{ \lam v_k}{\ome_1-\ome_k + i\eps} + O(\lam^3)
\EQN{cklam}
\eeqa
which leads to
\beqa
\Big|\frac{\lam  v_k }{(z-\ome_k)^+_{z_1}}\Big|^2 &=&   \frac{ \lam^2 v_k^2}{|\ome_1-\ome_k + i\eps|^2} + O(\lam^4)
 \EQN{ck2}\\
 &=& {\pi\over\eps} \lam^2 v_k^2 \del(\ome_1-\ome_k)+ O(\lam^4) \to\infty.
\nonumber
\eeqa
This diverges when $\eps\to 0$. Hence \Ep{ck2'} is nonanalytic at $\lam=0$ due to the resonance at $\ome_1=\ome_k$. We have Poincar\'e divergence in the perturbation series of $(\Lam^{\dagger}q^*_1)(\Lam^{\dagger}q_1)$.

We note that when the energy of the field is non-extensive, we
have $\bra J_k\ket \sim O(1/L)$. The energy density goes to zero
in the infinite volume limit. In this case the appearance of the
Poincar\'e divergence in \Ep{ck2'} has no effect on the particle.

For quantum mechanics the situation is different. We can have fluctuations even in non-extensive situations \cite{OPP} due to vacuum effects. For example we obtain, for a two-level atom, an energy fluctuation of the dressed excited state which is of the order of the decay rate. This gives an uncertainty relation between energy and lifetime.

Coming back to our main discussion, we conclude that $\Lam^\dagger q^*_1 q_1$ cannot be expressed as the product \Ep{Q*1Q1} since $\Lam$ is, by definition, analytic in the coupling constant. To make this transformed product analytic, we make the replacement
\beqa
 \lam^2 \sum_k |c_k|^2 q_k^* q_k \Rightarrow \lam^2 \sum_k \xi_k q_k^* q_k
 \EQN{analrep}
\eeqa
where $\xi_k$ is a suitable analytic function. Due to the requirements on $\Lam$ stated in the Introduction this function is not quite arbitrary. Indeed, in the integrable case the term $q_k^* q_k$ would appear in $U^\dagger q_1^* q_1$ as $ \lam^2 \sum_k \cb_k^2 q_k^* q_k$ (see \Ep{Qb_1-6}). In the nonintegrable case $\cb_k$ is extended to $c_k$ or $c_k^*$, and becomes complex. Taking into account the requirements (1), (3) and (4) in the Introduction we conclude that a suitable extension of $\cb_k^2$ to the nonintegrable case is the linear superposition
\beqa
 \xi_k = rc_k^2 + {\rm c.c.},  \quad r+r^*=1
 \EQN{xikdef}
\eeqa
where $r$ is a complex constant to be determined. The relation $r+r^*=1$ is the simplest relation which guarantees  $\xi_k$ reduces to $\cb_k^2$ in the integrable case [see also the comments below \Ep{LamGam}].

 So we have \footnote{ Neglecting $O(1/L)$ terms, the second term in \Ep{Lam-J1} may be expressed in terms of renormalized field modes as $\sum_k b_k \Qt^*_k \Qt_k$ (see \Ep{qtk-sec8}). }
\beqa
\Lam^{\dagger}q^*_1 q_1 = \Qt^*_1 \Qt_1 + \sum_k b_k q^*_k q_k
 \EQN{Lam-J1}
\eeqa
where
\beqa
 b_k = \lam^2 |N_1|(-|c_k|^2 + \xi_k).
 \EQN{bkdef}
\eeqa

As shown in Appendix \ref{app:r} using the requirement (2) we obtain
\beqa
r = \frac{\exp{(-ia/2)}}{2 \cos (a/2)}, \quad N_1  = |N_1|\exp{(-ia)}
 \EQN{rdef}
\eeqa
giving a concrete form of $\Lam$ in \Ep{Lam-J1}.
By including the term $b_k$ in \Ep{Lam-J1} we have removed the Poincar\'e divergence in the product of Gamow modes. As a consequence,
\beqa \Lamd q_1^* q_1  \ne (\Lamd q_1^*)(\Lamd q_1).
 \EQN{Lq1Lq1}
\eeqa
This shows the non-distributive property of $\Lam$.

For weak coupling the approximate value of $b_k$ is given by \cite{OPP},
\beqa
b_k \approx \frac{2\pi}{L} \frac{\lambda^2 v_k^2 \gamma^2}{[(\ome_k -\omet_1)^2 + \gamma^2]^2}. \EQN{bk-secfp}
\eeqa
This has a sharp peak at $\ome_k = \omet_1$ with a width $\gamma$. It corresponds to the line shape of emission and absorption of the field by the renormalized particle.

 To find more general transformed products $\Lam^{\dagger}q^{*m}_1 q_1^n$, we apply the same logic that led to \Ep{Lam-J1}. Whenever $|c_k|^2$ appears in $\Qt^{*m}_1 \Qt^{n}_1 $, we replace it with $\xi_k$. This leads to (see  Appendix \ref{apx:analytic}).
\beqa
&&\Lam^{\dagger} q^{*m}_1 q^n_1 \EQN{LAM-MAIN} \\
&&= \sum_{a=0}^{min(m,n)} \frac{m!n!}{(m-a)!(n-a)!a!}\Qt^{*m-a}_1\Qt^{n-a}_1 Y^a
\nonumber
\eeqa
where $min(m,n)$ is the smaller of $m$, $n$, and
\beqa Y \equiv \sum_k b_k q_k^*q_k.
 \EQN{Ydef} \eeqa %%
Note that $b_k \sim O(1/L)$. Hence $Y \sim O(L^0)$ only if the field obeys the thermodynamic limit condition, \Ep{thlim}.
Otherwise $Y$ vanishes as $1/L$ and $\Lamd$ becomes distributive.  Also, when there are no resonances, $z_1$ becomes real and both $b_k$ and $Y$ vanish. Then $\Lamd$ reduces to $U^\dagger$ (see \Ep{Uqmn}).  In short, both  thermodynamic limit and resonances are necessary to obtain non distributivity of $\Lamd$ in \Ep{LAM-MAIN}, which, as we will see in the next section, gives the fluctuations found in Brownian motion.

For $\Lam^{-1} q^{*m}_1 q_1^n$ we obtain the expression (\ref{LAM-MAIN}) with $\Qt_1, \Qt_1^*$ replaced by $Q_1, Q_1^*$, respectively.

\subsection{Obtaining closed Markovian equations}
%%%%%%%%%%%%%%%%%%%%%%%%%%%%%%%%%%%
The $\Lam$ transformation we have presented satisfies all our requirements
(1)-(4) stated in the Introduction.  Now we show that $\Lam$ also satisfies the requirement (5) namely, that $\Lam$ gives closed Markovian equations.  To obtain closed Markovian kinetic equations, we first operate $\Lam$ on the Liouville equation, to obtain \Ep{rhot-eq}.
Kinetic equations involve a projection (or a part) of the ensemble $\rhot$.
In order for $\Lam$ to give closed kinetic of equations, we require that
the transformed Liouvillian $\tht$ in \Ep{rhot-eq} leaves subspaces corresponding to projections of  $\rhot$ invariant.  We will represent these subspaces by projection operators $P^{(\nu)}$, which are orthogonal  and  complete in the domain of $\tht$,
\begin{equation}
    P^{(\mu)} P^{(\nu)} =  P^{(\mu)}\delta_{\mu\nu}, \qquad
    \sum_\nu P^{(\nu)} =1.
    \EQN{Pcomplete}
\end{equation}
The invariance property of $\tht$ is
\begin{equation}
 P^{(\nu)}   \tht  =  \tht P^{(\nu)}.
    \EQN{OP=PO}
\end{equation}
Thanks to this commutation property, we obtain from \Ep{rhot-eq} closed Markovian equations for the  projections of $\rhot$,
\beqa
 i\frac{\partial}{\partial t}P^{(\nu)} \rhot(t) = \tht P^{(\nu)} \rhot(t).
 \EQN{Prhot-eq}
\eeqa

We choose $P^{(\nu)}$ as eigenprojectors of $L_0$. We have $L_0 P^{(\nu)} = w^{(\nu)}P^{(\nu)}$ where $w^{(\nu)}$ are the eigenvalues.
Then, for the integrable case the relation (\ref{OP=PO}) is automatically satisfied, since
in this case $\tht$ reduces to ${\bar L}_0$, the renormalized free Liouvillian with eigenprojectors $P^{(\nu)}$.

In the Friedrichs model the $P^{(\nu)}$ subspaces consist of monomials (or superposition of monomials) of field and particle
modes. For example the monomials
\beqa
 q_1^* q_k = P^{(1k)} q_1^* q_k,   \quad q_1^* q_k q_l^*q_l =  P^{(1k)} q_1^* q_k q_l^*q_l
 \EQN{exPnu'}
\eeqa
belong to the same subspace $P^{(1k)}$ with eigenvalue $w^{(1k)} \equiv \ome_1-\ome_k$.

One may introduce a Hilbert space structure for the
eigenfunctions of $L_0$, including suitable normalization constants in  the Segal-Bargmann representation \cite{rdamping}.
We also note that $\Lamd$ transformed variables $\Lamd P^{(\nu)} A$ generate the kinetic equation (\ref{Prhot-eq}) as we have $\bbra  \Lamd P^{(\nu)} A| \rho(t)\kket = \bbra A|P^{(\nu)}|\rhot\kket$.

Now we verify that the relation (\ref{OP=PO}) is satisfied for the $\Lam$ transformation we have constructed. We restrict ourselves to the components associated with the monomials in \Ep{Lqmn}. These belong to eigenspaces  of $L_0$ with eigenvalues $(m-n)\ome_1$. We denote the corresponding projectors as    $P^{(mn)}$.
Using \Ep{3-moment-apx}  in Appendix \ref{apx:moments} with $q_1'=0$, we find
\begin{equation}
    \tht^\dagger q_1^{*m} q_1^n = [(mz_1^*-nz_1) q_1^* q_1 - 2i\gamma m n  Y] q_1^{*m-1} q_1^{n-1}
 \EQN{thdqq}
\end{equation}
and similarly
\begin{equation}
    \tht  q_1^{*m} q_1^n = [(mz_1-nz_1^*) q_1^* q_1 + 2i\gamma m n  Y] q_1^{*m-1}
    q_1^{n-1}.
 \EQN{thqq}
\end{equation}
Both  the l.h.s. and the  r.h.s. of these two equations belong to the same eigenspace $P^{(mn)}$. This illustrates the statement that $\tht$ leaves the subspaces $P^{(\nu)}$ invariant, satisfying the requirement (5).

Due to the  $Y$ term, $q_1^{*m} q_1^n$ are  not eigenfunctions of $\tht$, so $P^{(mn)}$ is not an eigenprojector of $\tht$. This is quite natural, since the kinetic processes include both the decay of the particle modes (through emission of the field) and the absorption of field modes. These correspond to the first and second terms  inside brackets in Eqs. (\ref{thdqq}), (\ref{thqq}), respectively.

\subsection{$\Pi$ subdynamics}
%%%%%%%%%%%%%%%%%%
We comment on the  theory of $\Pi$ subdynamics, developed by the Brussels school
\cite{Rosen}. We introduce the projectors
\begin{equation}
   \Pi^{(\nu)} = \Lam^{-1} P^{(\nu)} \Lam .
 \EQN{Pidef}
\end{equation}
Similar to \Ep{Pcomplete}, they are orthogonal and complete. From \Ep{OP=PO} it follows that $\Pi^{(\nu)} L_H = L_H \Pi^{(\nu)}$.
Hence these projectors define independent subspaces that follow independent, closed dynamics. The projectors themselves may be written in terms of generalized eigenstates of $L_H$, which give a complex spectral representation of this operator \cite{PP97}.

 From the completeness relation of the $P^{(\nu)}$ or $\Pi^{(\nu)}$ projectors, we can recover the time evolution in the original variables as
\beqa \rho(t) = \sum_\nu \Lam^{-1} P^{(\nu)} \rhot(t) =  \sum_\nu
\Pi^{(\nu)} \rho(t).
 \EQN{recover}
\eeqa
This shows that, as pointed out in the Introduction, dynamics is decomposed into a set of components obeying Markovian equations. In order to actually use \Ep{recover} we need to know all the components $P^{(\nu)} \rhot(t)$ (or equivalently, $\Pi^{(\nu)} \rho(t)$). At the present moment we have only obtained a restricted set of these. This is enough for our present goal. Rather than solving the original equations of motion (which can be done by other standard methods) our goal is to show that Brownian motion, Gaussian white noise and damping are part of dynamics seen in the $\Lam$ representation.

The contruction of $\Lam$ we have presented here is based on renormalized particle modes. A more general construction of $\Lam$ starts with
 the commutation relation (\ref{OP=PO}) together with the other requirements stated in the Introduction. The main idea is to associate a ``degree of correlation'' with each subspace  $P^{(\nu)}$. Dynamics induces transitions among different $P^{(\nu)}$ subspaces. We have a ``dynamics of correlations'' \cite{IP62}. This allows us to perform the regularization of denominators of $U$ in a systematic way, depending on types of transitions (from lower to higher correlations or vice versa), which leads to $\Lam$.  The interested reader can find a presentation of this formulation  in  Refs. \cite{Rosen,OPP,rdamping}.

%%%%%%%%%%%%%%%%%%%%%%
% Section 6
%%%%%%%%%%%%%%%%%%%%%%
\section{Comparison with the phenomenological Langevin equation}
\label{sec:lang}

In this Section we discuss the relation between  the solution of the phenomenological Langevin equation for the Friedrichs model and $\Lamd$ transformed particle modes (i.e. dressed modes).  We will focus on the $\Lam^\dagger$ transformation, so that the transformed variables decay for $t>0$ \footnote{Note that variables $A$ evolve as $\exp(iL_H t) A$, while states $\rho$ evolve as  $\exp(-iL_H t) \rho$. In Refs. \cite{OPP,rdamping,POP2001} we considered transformed {\it states} that decay for $t>0$.  For this reason in those papers we used the $\Lam^{-1}$ transformation rather than $\Lamd$ (see Eqs. (\ref{LHA}), (\ref{ABL})).}  (see \Ep{At1dec}).  Remarkably, the Langevin equation and the dynamical equations for dressed modes have the same solution.

The phenomenological Langevin equation for the Brownian harmonic oscillator appropiate for the Friedrichs model has the form ($t>0$)
\beqa
\frac{d}{dt} \hq_1(t) = -i\hz_1\hq_1(t)+ \hR(t), \EQN{a1-sec-lang}
\eeqa
where $\hR(t)$ is a complex noise source (see \Ep{stand4}). We use the hats to denote phenomenological variables. The complex coefficient $\hz_1 \equiv \home_1 -i\hgam$ gives the frequency  and the damping rate of the oscillator. We assume $\hR(t)$ has the following properties:
\vskip .5cm
 (1) $\hR^*(t)$ and $\hR(t')$ have the delta function (white noise) correlation.
\beqa
\bra \hR^*(t)\hR(t') \ket =  \hR_c^2 \delta (t-t') \EQN{90}
\eeqa
where $\bra\, \ket$ means noise average.

(2) $\hR(t)$ has the Gaussian property
\beqa
& &\bra \hR^*(t_1)...\hR^*(t_m)\hR(t'_1)...\hR(t'_n) \ket \EQN{R2-sec-lang} \\
& &= \delta_{mn} \sum_{\mbox{all pairs}} \bra \hR^*(t_{i_1})\hR(t'_{j_1})\ket\cdots \bra  \hR^*(t_{i_n})\hR(t'_{j_n})\ket. \nonumber
\eeqa
The noise constant $\hR_c$ is determined in Appendix \ref{apx:AcBc} using equipartition of energy and assuming the bath is at temperature $T$. The result is $\hR_c^2= 2\hgam k_B T/\home_1$.

\Ep{a1-sec-lang}  corresponds to the equations
\beqa
\frac{d}{dt}\hx_1(t) &=& -\hgam \hx_1(t) + \frac{\hp_1(t)}{\hat m} + A(t), \EQN{01-sec-lang}\\
\frac{d}{dt}\hp_1(t) &=& -\hgam \hp_1(t) - {\hat m} \home_1^2 \hx_1(t) + B(t). \EQN{02-sec-lang}
\eeqa
where $A(t)$ and $B(t)$ are independent Gaussian white noises \cite{Sungyun}. These equations describe a damped harmonic oscillator with random momentum and force terms $A(t)$ and $B(t)$, respectively.   The equations  are symmetrical under rescaled position and momentum exchange, which is consistent with the same symmetry of the Hamiltonian.

The solution of \Ep{a1-sec-lang} is given by
\beqa
\hq_1(t) = \hq_{1a}(t) + \hq_{1r}(t), \EQN{a1t-sec-lang}
\eeqa
where
\beqa
&&\hq_{1a}(t) \equiv \hq_1(0) e^{-i\hz_1 t}, \EQN{94'}\\
&&  \hq_{1r}(t) \equiv e^{-i\hz_1 t} \int_0^t dt'\hR(t') e^{i\hz_1 t'} \EQN{94''}.
\eeqa
The term $\hq_{1a}(t)$ describes the damped harmonic oscillator without noise, and the term $\hq_{1r}(t)$ describes the behavior due to the noise.

For later comparison we calculate the autocorrelation function $\bra \hq_1^{*m}(t) \hq_1^n (t) \ket $.  We have
\beqa
& &\bra \hq^{*m}_q(t) \hq^n_1(t) \ket \nonumber \\
& &= \bra (\hq^*_{1a}(t) + \hq^*_{1r}(t))^m (\hq_{1a}(t)+\hq_{1r}(t))^n \ket \nonumber \\
& &= \sum_{k=0}^m\sum_{l=0}^n  \frac{m!}{(m-k)! k!} \frac{n!}{(n-l)!l!} \nonumber\\
& &\times  \hq_{1a}^{*m-k}(t)\hq_{1a}^{n-l}(t) \bra \hq_{1r}^{*k}(t) \hq_{1r}^l (t) \ket. \EQN{amn-sec-lang}
\eeqa
 The quantity  $\bra \hq_{1r}^{*k}(t) \hq_{1r}^l (t) \ket$ is non-zero only when $k=l$, as we can see from \Ep{R2-sec-lang}. Considering the fact that the number of sets of all possible pairs in $\bra \hR^*(t_1)...\hR^*(t_l)\hR(t_1')...\hR(t_l') \ket $ is $l!$, we have
\beqa
& &\bra \hq_{1r}^{*k}(t) \hq_{1r}^l(t) \ket \nonumber \\
& &= l! \delta_{kl} \bigg ( \bra e^{i\hz_1^* t} \int_0^t dt_1 \hR^*(t_1) e^{-i\hz_1^* t_1} \right .\nonumber\\
& &\times\left . e^{-i\hz_1 t}\int_0^t dt_2 \hR(t_2) e^{i\hz_1 t_2} \ket \bigg )^l \nonumber \\
& &=  l! \delta_{kl} \bigg ( \frac{\hR_c^2 (1-e^{-2\hgam t})}{2\hgam} \bigg )^l \nonumber\\
& &= l! \delta_{kl}   \Big(\frac{k_B T}{\home_1}\Big)^l
(1-e^{-2\hgam t})^l .
\EQN{arr-sec-lang} \eeqa %%
Substituting \Ep{94'} and \Ep{arr-sec-lang} into \Ep{amn-sec-lang}, we get
\beqa
& &\bra \hq_1^{*m}(t) \hq_1^n(t) \ket \nonumber \\
& &= e^{i(m\hz_1^*-n\hz_1)t} \sum_{l=0}^{min(m,n)} \frac{m!n!}{(m-l)!(n-l)!l!} \nonumber \\
& &\times \hq_1^{*m-l}(0) \hq_1^{n-l}(0) \Big(\frac{k_B T}{\home_1}\Big)^l(e^{2\hgam t}-1)^l.
\EQN{avg-sec-lang}
\eeqa

Now we can compare the above expression with the time-evolved dressed products $e^{iL_H t} \Lamd q_1^{*m}q_1^n$. We have (see \Ep{LHAt} and \Ep{LAM-MAIN})
\beqa
&&e^{iL_H t} \Lamd q_1^{*m}q_1^n = \sum_{a=0}^n \frac{m!n! }{(m-a)!(n-a)!a!}  \nonumber \\
&&\times e^{i(mz_1^* -nz_1)t}e^{2\gam a t}  \Qt^{*m-a}\Qt^{n-a}
Y^{a} .
 \EQN{Lam-sec-lang0} \eeqa %%
Writing
\beqa
 e^{2\gam a t}  = \sum_{l=0}^a \frac{a!}{l!(a-l)!} (e^{2\gam t}-1)^l
\EQN{Lam-sec-lang1}
\eeqa
and $l'=a-l$ we have
\beqa
&&e^{iL_H t} \Lamd q_1^{*m}q_1^n = \sum_{l=0}^n \sum_{l'=0}^{n-l} \frac{m!n! }{(m-l-l')!(n-l-l')!l!l'!}\nonumber \\
&&\times e^{i(mz_1^* -nz_1)t}\Qt^{*m-l-l'}\Qt^{n-l-l'} Y^{l+l'}(e^{2\gam t}-1)^l \nonumber \\
&&= \sum_{l=0}^n  \frac{m!n!}{(m-l)!(n-l)!l!}e^{i(mz_1^*-nz_1)t} \nonumber \\
&&\times \sum_{l'=0}^{n-l}\frac{(m-l)!(n-l)!}{(m-l-l')!(n-l-l')!l'!} \Qt^{*m-l-l'}\Qt^{n-l-l'} Y^{l'} \nonumber \\
&&\times  Y^l (e^{2\gam t}-1)^l .
\EQN{Lam-sec-lang2} \eeqa %%
Using \Ep{LAM-MAIN} again we obtain
\beqa
&&e^{iL_H t} \Lamd q_1^{*m}q_1^n  \nonumber\\
&&= e^{i(mz_1^*-nz_1)t} \sum_{l=0}^n \frac{m!n!}{(m-l)!(n-l)!l!} \nonumber \\
&&\times (\Lamd q^{*m-l}_1q^{n-l}_1 ) Y^l (e^{2\gam t}-1)^l .
\EQN{Lam-sec-lang} \eeqa %%
Comparing \Ep{avg-sec-lang} and \Ep{Lam-sec-lang}, we see the direct correspondences
\beqa
\hz_1  &\Leftrightarrow& z_1 \nonumber\\
 \frac{k_B T}{\home_1}  &\Leftrightarrow& Y=\sum_k b_k q^*_k q_k \EQN{Y-sec-lang}\\
 \bra \hq_1^{*m}(t) \hq_1^n(t) \ket  &\Leftrightarrow& e^{iL_H t} \Lamd (q_1^{*m}q_1^{n}) \nonumber.
\eeqa
The form and time evolution of the ensemble average of Langevin equation variables are the same as those of $\Lam$ transformed variables.
 Furthermore, if we take the ensemble average of $\Lamd q_1^{*m}q_1^n$, we see a closer correspondence. Let us assume that the field action $J_k$ obeys the unperturbed Gibbs distribution. The initial distribution $\rhot_0(\Gamma)$ has the form (with $\beta \equiv1/(k_B T)$)
\beqa
\rhot_0 (\Gamma) = C \rho_{01}(J_1,\alp_1) \exp(-\beta \sum_k \ome_k J_k)
\eeqa
where $C$ is a normalization constant, $k_B$ is Boltzmann's constant and $T$ is the temperature. As shown in \Ep{rsing6}
the average of $J_k$ over this ensemble is
\beqa
\bra J_k \ket =  \frac{1}{\ome_k \beta} = \frac{k_B T}{\ome_k}. \EQN{Jkavg-sec-fp}
\eeqa
To calculate $\sum_k b_k \bra J_k \ket $, we need the form of $b_k$.
 The approximate value of $b_k$ is given in \Ep{bk-secfp}, which for the weak coupling case is approximated by the delta function $(2\pi/L) \delta (\ome_k-\omet_1)$ \cite{OPP}. So we get
\beqa
\sum_k b_k \bra J_k \ket = \sum_k b_k \frac{k_B T}{\ome_k} \approx \frac{k_B T}{\omet_1}. \EQN{kbt-sec-fp}
\eeqa
Note that ${\ome_k}^{-1}$ does not make any divergence for small $k$ since $b_k$ is proportional to $v_k^2 \sim \ome_k$ for small $k$.

In short, we obtain a complete correspondence between $\Lam$ transformed modes and Langevin modes  (see \Ep{Y-sec-lang}).  The systematic removal of Poincar\'e divergences in the $\Lam$ transformation gives the Gaussian white noise structure.

%%%%%%%%%%%%%%%%%%%%%%
% Section
%%%%%%%%%%%%%%%%%%%%%%
\section{The Fokker-Planck equation}
\label{sec:fp-drv}

Using the above results we can now derive the Fokker-Planck equation for the transformed density function $\rhot = \Lam\rho$. We start with the transformed equation (see \Ep{rtdef})
\beqa
 i\frac{\partial}{\partial t}\rhot = \tht \rhot. \EQN{eq-tht}
\eeqa
 We derive the Fokker-Planck equation for $q_1$, $q^*_1$. We follow the standard derivation found in
 textbooks  (see \cite{Balescu, Resibois}), but now in terms of $\Lam$. Consider a test function $G(q_1, q^*_1)$,
 which is smooth and vanishes at $|q_1| = \infty$.  Multiplying this on both sides of \Ep{eq-tht} and integrating over the phase space, we have
\beqa
& &\int d\Gam G(q_1, q^*_1)i\frac{\partial}{\partial t}\rhot(\Gam, t) = \int d\Gam G(q_1, q_1^*) \tht(\Gam) \rhot(\Gam, t) \nonumber \\
& &= \int d\Gam d\Gam' G(q_1,q^*_1) \tht(\Gam) \delta(\Gam-\Gam') \rhot(\Gam', t). \EQN{basic}
\eeqa
 In \Ep{basic}, $\tht (\Gam)$ means $\tht$ that acts on $\Gam$ variables.
 We expand $G(q_1,q^*_1)$ near $q_1'$ and $q^{'*}_1$.
\beqa
&&\int d\Gam G(q_1, q_1^*)i\frac{\partial}{\partial t}\rhot(\Gam, t) \EQN{basic2} \\
&&=  \int d\Gam d\Gam' \bigg \{ \sum_{m=0}^{\infty}\sum_{n=0}^{\infty}
 \frac{1}{m!n!} \bigg(\frac{\partial^m}{\partial (q_1^{'*})^m}\frac{\partial^n}{\partial (q_1')^n} G(q_1', q_1^{'*}) \bigg)
\nonumber\\
&&\times (q_1^*-q_1^{'*})^m (q_1-q_1')^n \bigg \}
 \tht (\Gam)  \delta(\Gam-\Gam') \rhot(\Gam', t). \nonumber
\eeqa
Integrating by parts, \Ep{basic2} becomes
\beqa
& &\int d\Gam G(q_1, q_1^*)i\frac{\partial}{\partial t}\rhot(\Gam, t) \EQN{pre-fp} \\
& &=  \int d\Gam' G(q_1',q_1^{'*}) \sum_{m=0}^{\infty}\sum_{n=0}^{\infty}\frac{(-1)^{m+n} }{m!n!}\frac{\partial^m}{\partial (q_1^{'*})^m}\frac{\partial^n}{\partial (q_1')^n}
\nonumber \\
& &\times  \bigg [\int d\Gam (q_1^*-q_1^{'*})^m (q_1-q_1')^n  \tht
(\Gam)  \delta(\Gam-\Gam') \bigg ] \rhot(\Gam', t) .\nonumber
\eeqa %%
We call the quantities inside the brackets in \Ep{pre-fp} the ``moments'' of order $m+n$. The moments are calculated explicitly in Appendix~\ref{apx:moments}. They are given by
\beqa
&&\int d\Gam (q_1^*-q_1^{'*})^m (q_1-q_1')^n \tht (\Gam) \delta (\Gam-\Gam')  \nonumber\\
&&= \left\{ \begin{array}{c}
                      z_1q_1', \quad m=0,\,n=1\\
                      -z_1^*q^{'*}_1,\quad m=1,\,n=0 \\
                        2i\gam \sum_k b_k q^{'*}_k q_k',\quad m=1,\,n=1 \\
                        0,\quad \mbox{for all other $m$ and $n$.}
\end{array} \right. \EQN{moments}
\eeqa
Substituting \Ep{moments} into \Ep{pre-fp}, we get
\beqa
& & \int d\Gam G(q_1,q^*_1) i \frac{\partial}{\partial t} \rhot(\Gam,t)\EQN{int-fp} \\
& &= \int d\Gam' G(q_1',q^{'*}_1)(-\frac{\partial}{\partial q_1'})(z_1 q_1')\,\rhot(\Gam',t)  \nonumber \\
& &+ \int d\Gam' G(q_1',q^{'*}_1) (-\frac{\partial}{\partial q^{'*}_1})(-z_1^* q^{'*}_1)\,\rhot(\Gam',t) \nonumber \\
& &+ \int d\Gam' G(q_1',q^{'*}_1) (\frac{\partial^2}{\partial
q_1'\partial q^{'*}_1}) (2i\gam \sum_k b_k q^{'*}_k q_k' )\,
\rhot(\Gam',t). \nonumber \eeqa %%

 By changing the integration variable $\Gam'$ to $\Gam$ in the right hand side of \Ep{int-fp} and eliminating $i$ on both sides, we have
\beqa
& &\int d\Gam G(q_1,q^*_1) \frac{\partial}{\partial t}\rhot(\Gam,t) \nonumber\\
&=& \int d\Gam  G(q_1,q^*_1)\bigg \{ \frac{\partial}{\partial q_1}(iz_1 q_1) + \frac{\partial}{\partial q^*_1} (-iz_1^* q_1^*)\nonumber \\
&+&\frac{\partial}{\partial q_1 \partial q^*_1 } (2\gam\sum_k b_k q^*_kq_k ) \bigg\} \rhot(\Gam,t). \EQN{fac1-sec-fp}
\eeqa
Now suppose that $\rhot(\Gam,t)$ is factorized at $t=0$. In other words, we write
$\rhot(\Gam,0)$ as
\beqa
\rhot(\Gam,0) = g_1(q_1,q_1^*) \prod_{k}g_k(q_k,q_k^*). \EQN{fac}
\eeqa
As shown in Appendix~\ref{apx:fac}, this factorized form of $\rhot$ enables us to write \Ep{fac1-sec-fp} as
\beqa
& &\int d\Gam_1 G(q_1,q^*_1) \frac{\partial}{\partial t} \int d\Gam_f \rhot(\Gam,t) \nonumber\\
&=& \int d\Gam_1  G(q_1,q^*_1)\{ \frac{\partial}{\partial q_1}(iz_1 q_1)
+ \frac{\partial}{\partial q^*_1} (-iz_1^* q_1^*) \nonumber \\
&+& \frac{\partial^2}{\partial q_1 \partial q^*_1} (2\gam\sum_k b_k \bra q^*_kq_k \ket )\int d\Gam_f  \rhot(\Gam,t).
\EQN{fac-sec-fp}
\eeqa
Since $G(\Gam_1)$ is an arbitrary test function, we can write \Ep{fac-sec-fp} as
\beqa
 \frac{\partial}{\partial t} \rhot_1(\Gam_1,t) &=&   \Big\{ iz_1 \frac{\partial}{\partial q_1} q_1 -iz_1^* \frac{\partial}{\partial q^*_1} q^*_1
\EQN{int-fp-fac2} \\
 &+& 2\gam\sum_k b_k \bra J_k \ket \frac{\partial}{\partial q_1 \partial q^*_1} \Big\}\  \rhot_1(\Gam_1,t),
\nonumber
\eeqa
 where
\beqa
\rhot_1(\Gam_1,t) = \int d\Gam_f \rhot(\Gam,t).
\eeqa
\Ep{int-fp-fac2} is our   Fokker-Planck equation for the normal modes.
 This equation is applicable for any initial field configuration obeying
 the extensive condition and \Ep{fac}. In the non-extensive case, the diffusion term containing $b_k$ vanishes, and the equation describes damping of the oscillator without Brownian motion. For the special case where the field has the unperturbed Gibbs distribution, using the approximation (\ref{kbt-sec-fp})  we get
\beqa
 \frac{\partial}{\partial t} \rhot_1(\Gam_1,t) &\approx&   \{ iz_1\frac{\partial}{\partial q_1}q_1 -iz_1^*  \frac{\partial}{\partial q^*_1} q_1^*\nonumber \\
&+& \frac{2\gam k_B T}{\omet_1} \frac{\partial^2}{\partial q_1 \partial q^*_1}\}\  \rhot_1(\Gam_1,t).
\EQN{int-fp-fac3}
\eeqa
This is precisely the equation one obtains from the phenomenological Langevin equation (\ref{a1-sec-lang}).

 The Fokker-Planck equation for other variables can be also derived from \Ep{int-fp-fac2} by changing variables.  For example, the Fokker-Planck equation for the position and momentum $x_1$ and
$p_1$ is given by
\beqa
\frac{\partial}{\partial t} \rhot_1(\Gam_1,t) &=&   \Big\{ -\frac{\partial}{\partial x_1}(\frac{p_1}{\mt}-\gam x_1) + \frac{\partial}{\partial p_1} (\mt\omet_1^2 x_1 +\gam p_1)\nonumber \\ &+&\frac{D_x}{2}\frac{\partial^2}{\partial x_1^2} + \frac{D_p}{2}\frac{\partial^2}{\partial p_1^2} \Big\}\  \rhot_1(\Gam_1,t),
\EQN{int-fp-fac4}
\eeqa
where
\beqa
\mt &=& m\ome_1 /\omet_1, \nonumber\\
D_x &=&  \frac{2\gam}{\mt \omet_1} \sum_k b_k \bra J_k \ket \approx   \frac{2\gam k_B T}{\mt \omet_1^2},\nonumber\\
 D_p &=&  2\mt \gam \omet_1 \sum_k b_k \bra J_k \ket \approx 2\mt \gam k_B T.
\eeqa
[the approximate values are applicable for the unperturbed Gibbs distribution.]
 The Fokker-Planck equation for the action variable $J_1$ is given  (after integration over the angle variable $\alp_1$) by
\beqa
&&\frac{\partial}{\partial t} \rhot (J_1, t)  \EQN{FokkJ}\\
&& = \Big\{ 2\gam\frac{\partial}{\partial J_1} ( J_1 -\frac{k_B T}{\omet_1}) + D_J \frac{\partial^2}{\partial J_1^2}
J_1  \Big\} \rhot (J_1, t),
\nonumber
\eeqa
where
\beqa
D_J = 2\gam  \sum_k b_k \bra J_k \ket  \approx \frac{2\gam  k_B T}{\omet_1}.
\eeqa

Eqs. (\ref{int-fp-fac4}) and (\ref{FokkJ}) coincide (in the weak-coupling approximation) with the equations for Brownian motion
of an oscillator in an anharmonic lattice derived in Ref. \cite{IP62}. \Ep{FokkJ} (in its exact form) was first proposed by T. Petrosky \cite{ppri}.

Note that  \Ep{int-fp-fac4} is symmetric with respect to rescaled position $x_1$ and momentum $p_1$. The reason  is that  the Hamiltonian considered here is symmetric in rescaled $x_1$ and $p_1$  to begin with. The same is true for the anharmonic lattice model considered  in Ref. \cite{IP62}.  In contrast, the Kramers (Fokker-Planck) equation \cite{IP62,vanKampen} derived from the Ornstein-Uhlenbeck phenomenological theory of Brownian motion \cite{Brownian} is not symmetric, because the Brownian force breaks the position-momentum symmetry. In spite of the difference,  for the case $\gamma\ll\ome_1$, \Ep{int-fp-fac4} gives the same solution as the Kramers equation. The solutions of \Ep{int-fp-fac4} can be found in Ref. \cite{IP62}.

%%%%%%%%%%%%%%%%%%
% Conclusions
%%%%%%%%%%%%%%%%%%
\section{Conclusions}

 In this paper we studied the irreversible and stochastic behavior of an oscillator coupled to a field in the thermodynamic limit, using the star-unitary transformation $\Lam$.  We showed that the average of dressed particle modes has the same time evolution as the ensemble average of particle modes in the Langevin equation (\ref{a1-sec-lang}).
Also, the reduced distribution function for the particle variables exactly obeys the Fokker-Planck equation (\ref{int-fp-fac2}), which describes the damping and diffusion processes.
It is remarkable that the systematic removal of Poincar\'e divergences by analytic continuation leads to the same structure as that of Gaussian white noise.

Since the Gaussian structure of the fluctuations is coming from the resonances, rather than the specific form of the initial field ensemble, our derivation of the Fokker-Planck equation is valid
for both Gaussian and non-Gaussian field ensembles. Due to the Fokker-Planck evolution, the particle distribution is still Gaussian for $t\to\infty$. 

Our method of isolating Poincar\'e divergences can in principle be applied  to more complicated systems than the one considered here, in order to investigate further the relation between noise and dynamics.

We have studied the fluctuations of a particle surrounded by a field. One can also consider the fluctuations
of the field modes induced by the particle. An interesting result is that fluctuations of the $\Lam$-transformed field modes appear even when the energy of the field is non-extensive. If we go back to the initial formulation, acting with $\Lam^{-1}$, we recover the usual equations. This will be discussed elsewhere \cite{rdamping}.

Markovian equations such as the Langevin or Fokker-Planck equations describe irreversible and stochastic processes. Rather than viewing these equations as approximations of Hamiltonian dynamics, we see them as describing components of dynamics. These components can be identified through a change of phase-space variables obtained by the $\Lambda$ transformation. This gives a representation with broken time-symmetry.

In the non-Markovian equations, the effects of dressing are  not separated from irreversible (or thermodynamic) processes. In our approach
the dressing on the  particle is incorporated from the beginning, since we deal with renormalized particles. This  allows us to isolate pure thermodynamic behavior such as Brownian motion with a white noise source. In this sense, we can aim to reformulate thermodynamics in terms of renormalized particles. This is interesting in view of recent claims \cite{Nieu}  that
traditional thermodynamics may not apply for quantum Brownian motion at low temperatures, due to  the non-Markovian character of
quantum noise.

The relation with thermodynamics requires a precise definition of entropy or ${\cal H}$ function. The latter can be given in terms of $\Lam$ \cite{POP2001}.  The introduction of $\Lam$ leads to probabilistic considerations in classical mechanics, independent of incomplete knowledge or quantum corrections.  In conclusion, we believe that our approach leads to a unification of dynamics, thermodynamics and noise.

%%%%%%%%%%%%%%%%%%%%
% Acknowledgments  %
%%%%%%%%%%%%%%%%%%%%
\acknowledgments

 We  thank Prof. I. Prigogine and Dr. T. Petrosky for their support and encouragement, as well as for many helpful comments and suggestions.  We thank the referee that reviewed our paper, for many  comments that have helped us to clarify our ideas and improve our presentation. We thank also Dr. G. Akguc, Prof. I. Antoniou, Prof. R. Balescu, Dr. D. Driebe, Prof. M. de Haan, Dr. E. Karpov, Mr. C. B. Li, Prof. G. Nicolis,  Prof. E.C.G. Sudarshan, Prof. S. Tasaki, and Mr. C. Ting for helpful comments. We acknowledge the International Solvay Institutes for Physics and
Chemistry, the Engineering Research Program of the Office of Basic Energy
Sciences at  the U.S. Department of Energy, Grant No DE-FG03-94ER14465, the
Robert A. Welch Foundation Grant F-0365, The European Commission Project HPHA-CT-2001-40002, the National Lottery of Belgium and the Communaut\'e Fran{\c}aise de Belgique for supporting  this work.

%%%%%%%%%%%%%%%%%%%%
% Appendices      %
%%%%%%%%%%%%%%%%%%%%

 \appendix

%%%%%%%%%%%%
%Appendix
%%%%%%%%%%%%%
\section{Exact time evolution of particle and field modes}
\label{apx:frs}

We consider first the integrable case. The time evolution of the modes $q_s(t)$ can be calculated using the renormalized modes $\Qb_s$, which are eigenstates of the Liouville operator. We have
\beqa
e^{iL_H t} \Qb_s =  e^{-i\omeb_s t} \Qb_s
\EQN{exact1}
\eeqa
The particle mode $\Qb_1$ is given \Ep{Qb_1-6}. The field modes $\Qb_k$ are found from the equation $L_H \Qb_k = -\omeb_k \Qb_k$, which gives
\begin{eqnarray}
   \Qb_k = N_k^{1/2} \Big[q_k +  {\lam V_k \over \eta_k(\omeb_k)}
  (q_1 + \sum_{k'(\ne k)}  {\lam V_{k'} \over \omeb_k -\ome_{k'}}q_{k'})
   \Big]
\end{eqnarray}
where
\begin{equation}
   \eta_k(z) \equiv z - \ome_1
    - \sum_{k' (\ne k)} {\lam^2 V_{k'}^2 \over z - \ome_{k'}},
  \EQN{Bog5}
\end{equation}
\begin{equation}
    \omeb_k = \ome_k + {\lam^2 V_k^2 \over  \eta_k(\omeb_k)},
  \EQN{Bog6}
\end{equation}
\begin{eqnarray}
    N_k = \Big[1 +  {\lam^2 V^2_k \over \eta^2_k(\omeb_k)}
  (1 + \sum_{k'(\ne k)}  {\lam^2 V^2_{k'} \over (\omeb_k
  -\ome_{k'})^2})\Big]^{-1}.
 \nonumber \\
\end{eqnarray}
Note that $\omet_k = \ome_k + O(1/L)$ and also $N_k = 1 + O(1/L)$. As long
as $L$ is finite all the denominators are non-vanishing, and there are no resonances.

We write the linear relations between dressed and bare modes as
\begin{eqnarray}
\Qb_s =  \sum_r c_{sr} q_r, \quad q_s = \sum_r c_{rs}^* \Qb_r .
 \EQN{exact2}
\end{eqnarray}
Then, using \Ep{exact1} we get the coefficients $f_{sr}(t)$ in the equation $q_s(t) =  \sum_r f_{sr}(t) q_r(0)$ as
\begin{eqnarray}
f_{sr}(t) = \sum_{r'} c_{rr'} e^{-i \omeb_{r'} t} c_{r's}^* .
 \EQN{exact3}
\end{eqnarray}

In the nonintegrable case resonances appear (see Sec. \ref{sec:Gam}) and if we insist on keeping the renormalized modes as usual canonical variables, then the particle modes dissappear into the continuum of field modes (we can however introduce the Gamow modes of Sec. \ref{sec:Gam}, which are non-canonical). Keeping canonical modes, the Hamiltonian is  represented as \cite{Friedrichs,OPP, rdamping}
\begin{eqnarray}
H = \sum_k \ome_k \Qt_k^* \Qt_k
 \EQN{exact4}
\end{eqnarray}
in the continuous limit. The renormalized field modes are given by
\beqa
\Qt_k = q_k + \frac{\lam V_k}{\eta^\mp  (\ome_k) } \bigg[ q_1 + \sum_{k'}\frac{\lam V_{k'}}{\ome_k -\ome_{k'}  \mp i\epsilon} q_{k'} \bigg ]
\EQN{qtk-sec8}
\eeqa
where $\eps$ is a positive infinitesimal quantity. There are two branches, corresponding to analytical continuation to the lower or upper half planes of $\ome_k$.  For $t>0$, and given the initial condition \Ep{nfok2},  it is convenient to take the branch with $-i\eps$, since this will give exponential decay of the particle modes the positive $t$ direction. We get
\beqa
&&q_1(t) = \sum_k \frac{\lam V_k }{\eta^+ (\ome_k)} \Qt_k (t) = \sum_k  \frac{\lam V_k }{\eta^+ (\ome_k)} \Qt_{k}(0)  e^{-i\ome_k t} \nonumber \\
&&= \sum_k  \frac{\lam V_k }{\eta^+ (\ome_k)} q_{k}(0)e^{-i\ome_k t} + \sum_k \frac{\lam^2 V_k^2}{|\eta^+(\ome_k)|^2} q_1(0) e^{-i\ome_k t} \nonumber \\
&&+ \sum_k \frac{\lam^2 V_k^2}{|\eta^{+} (\ome_k)|^2} \sum_{k'}
\frac{\lam V_{k'}}{\ome_k -\ome_{k'} -i\epsilon} q_{k'}(0) e^{-i\ome_k t}, \EQN{1-sec-ori}
\eeqa
For $t<0$, we take the  branch with $+i\eps$.

%%%%%%%%%%%%
%Appendix
%%%%%%%%%%%%%
\section{Randomness in the field modes}
\label{apx:thnoise}
 We choose our initial condition with the form (\ref{nfok2}).   In classical mechanics $q_{10}\equiv q_1(0)$ can be determined exactly since
$q_{10}$ is a function of the initial position and momentum of the particle.
 For the modes $q_{k0}$ we need more care. Let us first write $q_{k0}$ in terms of action and angle variables,
\beqa
q_{k0} = \sqrt{J_{k0}}\, e^{-i\alp_{k0}}.
\eeqa
With the ensemble (\ref{nfok2}) we have
\beqa
\lim_{L\to\infty}  \bra J_{k0} \ket  \sim O(L^0)
\eeqa
in the thermodynamic limit. For example, for an unperturbed Gibbs thermal distribution of the field modes we have $\bra J_{k0} \ket  = k_B T/\ome_k$ (see \Ep{rsing6}).

For almost all phase points $\{J_{10},..J_{k0}.., \alp_{10},..,\alp_{k0},..\}$ out of the  ensemble, any two different angles $\alp_{k0}$ and $\alp_{k'0}$  have no correlation. In other words, the sequence of angles $\{\alp_{k_n 0} \}$ is completely random for almost all cases. This   property allows the existence of the thermodynamic limit \cite{PP2000}. Indeed,  if $\alp_{k0}$ is uniformly distributed over $[-\pi,\,\pi]$ and  the sequence of angles $\{\alp_{k_n 0} \}$ is completely random, then the term
\beqa
& &\sum_k \frac{\lam V_k}{\eta^+ (\ome_k)} q_{k0} e^{-i\ome_k t} \nonumber \\
& &= \sum_k  \frac{\lam V_k}{\eta^+ (\ome_k)} \sqrt{J_{k0}}\, e^{-i(\ome_k + \alp_{k0})t} \EQN{1st-sec-ori}
\eeqa
in \Ep{1-sec-ori} is $O(L^0)$. This is so because the summation is taken over a sequence complex numbers with random phases and hence it is proportional to the square root of the number of modes, which in turn is proportional to $L$.  Since $V_k \sim L^{-1/2}$,
\Ep{1st-sec-ori} is $O(L^0)$.  It shows a very irregular time evolution as the number of modes increases.

Note that if $\alp_{k0}$ was a smooth function of $k$, then for the first term of \Ep{1-sec-ori} we would have in the limit $L \rightarrow \infty$
\beqa
& &\sum_k \frac{\lam V_k }{\eta^+ (\ome_k)} q_{k0} \nonumber \\
& &= \sqrt{\frac{L}{2\pi}} \frac{2\pi}{L} \sum_k \frac{\lam v_k }{\eta^+{(\ome_k)} }\sqrt{J_{k0}} e^{-i\alp_{k0}}  \nonumber \\
& &\rightarrow  \sqrt{\frac{L}{2\pi}} \int dk  \frac{\lam v_k }{\eta^+{(\ome_k)} }\sqrt{J_{k0}}\, e^{-i\alp_{k0}}
\EQN{3-sec-ori}
\eeqa
and since the integral is $O(1)$, this expression would diverge as  $O(\sqrt{L})$.

%%%%%%%%%%%%
%Appendix
%%%%%%%%%%%%%
\section{Derivation of the non-Markovian Fokker-Planck equation}
\label{apx:nmfok}
From the Liouville equation we have
\beqa
&&\int d\Gam G(\Gam_1)\frac{\partial}{\partial t}\rho(\Gam, t) = -i \int d\Gam G(\Gam_1)  L_H
 \rho (\Gam, t) \nonumber\\
 && = \sum_{m=0}^\infty \sum_{n=0}^\infty G_{mn}  M_{mn}(t)
\EQN{nfok1}
\eeqa
where
\beqa M_{mn}(t) \equiv -i  \int d\Gam q_1^{*m} q_1^n L_H  \rho
(\Gam, t).
\EQN{Mmndef} \eeqa %%
Using \Ep{Hc} we have
\beqa M_{mn}(t) =  \int d\Gam  \left(  i L_H [q_1^*(t)]^m
[q_1(t)]^n\right) \rho (\Gam, 0).
 \EQN{Mmn1} \eeqa %%
Since $L_H$ is a differential operator, we have
\beqa
 i L_H [q_1^*(t)]^m  [q_1(t)]^n &=& m[q_1^*(t)]^{m-1}  [q_1(t)]^n iL_H q_1^*(t) \nonumber\\
 &+& n [q_1^*(t)]^m  [q_1(t)]^{n-1} iL_H q_1(t).\nonumber\\
\EQN{Mmn2}
\eeqa
From Eqs. (\ref{stand1}) and  (\ref{stand4}) we have
\beqa
&& i L_H q_1^*(t) =  i z_1^*(t) q_1^*(t) + R^*(t), \nonumber\\
&& i L_H q_1(t) = - i z_1(t) q_1(t) + R(t).
 \EQN{Mmn3} \eeqa %%
Inserting Eqs. (\ref{Mmn2}) and (\ref{Mmn3}) in \Ep{Mmn1}  we get
\beqa
M_{mn}(t) =  M_{mn}^z(t) + M_{mn}^R(t)
\EQN{Mmn4}
\eeqa
where
\beqa
&&M_{mn}^z(t) \EQN{Mmn5}\\
&&=  \int d\Gam [i z_1^*(t) m  - iz_1(t) n] [q_1^*(t)]^m  [q_1(t)]^n  \rho (\Gam, 0)\nonumber\\
&&=  \int d\Gam [i z_1^*(t) m  - iz_1(t) n] q_1^{*m}  q_1^n  \rho (\Gam, t)\nonumber
\eeqa
and
\beqa
&&M_{mn}^R(t) = M_{mn}^{R,1}(t) + M_{mn}^{R,2}(t),\EQN{Mmn6}\\
&&M_{mn}^{R,1}(t) =  \int d\Gam\,  m  [q_1^*(t)]^{m-1}  [q_1(t)]^n R^*(t) \rho (\Gam, 0),\nonumber\\
&&M_{mn}^{R,2}(t) = \int d\Gam\,  n [q_1^*(t)]^m  [q_1(t)]^{n-1}
R(t)  \rho (\Gam, 0).
\nonumber \eeqa %%
Now we evaluate  $M_{mn}^{R,1}(t)$ in \Ep{Mmn6}.  Using  \Ep{stand5'} we have
\beqa
&&M_{mn}^{R,1}(t)\EQN{Mmn7}\\
&&=  \int d\Gam\,  m  [q_1^*(t)]^{m-1}  [q_1(t)]^n \sum_k h^*_k(t)
q_k^*\rho (\Gam, 0).\nonumber\eeqa %%
Then, using \Ep{sol1} we obtain
\beqa
M_{mn}^{R,1}(t)
&=& \int d\Gam\, m \left[f_{11}^*(t) q_1^* + \sum_{p} f_{1p}^*(t) q_p^*\right]^{m-1} \nonumber\\
&\times& \left[f_{11}(t) q_1 + \sum_{p'} f_{1p'} (t)q_{p'}\right]^n \EQN{mn10}  \\
&\times&  \sum_k h^*_k(t) q_k^* \rho(\Gam, 0). \nonumber \eeqa %%
Abbreviating $f_\alp\equiv f_{1\alp}(t)$, $h_k(t)\equiv h_k$ and using binomial expansions, we get
\beqa
&&M_{mn}^{R,1}(t) =  \int d\Gam\ \sum_{a=0}^{m-1} \sum_{b=0}^{n}
\frac{(m-1)!}{(m-1-a)!a!}  \frac{n!}{(n-b)!b!} \nonumber\\
&&\times
 m\,  [f_1^* q_1^*]^{m-1-a} [f_1 q_1]^{n-b}  \EQN{mn10a}\\
&&\times
\sum_{p_1\cdots p_{a+1}} \sum_{p_1'\cdots p_b'}  f^*_{p_1}\cdots f^*_{p_a}  h^*_{p_{a+1}}
 f_{p_1'}  \cdots f_{p_b'}
\nonumber\\
&&\times q^*_{p_1} \cdots q^*_{p_a} q^*_{p_{a+1}} q_{p_1'}\cdots  q_{p_b'}\,
 \rho(\Gam, 0) \nonumber
\eeqa
where we changed the variable $k$ to $p_{a+1}$
Due to the assumed form of the field distribution in \Ep{nfok2}, only expectation values of observables independent of the angles $\alp_p$ of the field modes are nonzero (in other words, each $q_{p_j'}$ field mode must be multiplied by its complex conjugate $q_{p_i}^*$ with $p_j'=p_i$). Thus in \Ep{mn10a} we must have $b=a+1$ and we get $(a+1)!$ possible parings
\beqa
&&M_{mn}^{R,1}(t) =  \int d\Gam\ \sum_{a=0}^{min(m-1,n-1)}
\nonumber\\
&& \times \frac{(m-1)!}{(m-1-a)!a!}  \frac{n!}{(n-1-a)!(a+1)!} \nonumber\\
&&\times
 m\,  [f_1^* q_1^*]^{m-1-a} [f_1 q_1]^{n-1-a}  \EQN{mn10b}\\
&&\times
\sum_{p_1\cdots p_{a+1}}  |f_{p_1}|^2 \cdots | f_{p_a}|^2  h^*_{p_{a+1}}  f_{p_{a+1}}
 \nonumber\\
&&\times (a+1)!\, q^*_{p_1} q_{p_1}\cdots q^*_{p_a} q_{p_a}
q^*_{p_{a+1}} q^*_{p_{a+1}}
 \rho(\Gam, 0). \nonumber
\eeqa
Note that due to the volume dependence of the coefficients $f_{p} \sim O(1/\sqrt{L})$ and
$h_p \sim 1/\sqrt{L}$ [which follows from \Ep{1-sec-ori}], we can neglect the cases where two
or more of the wave numbers are repeated, e.g., $p_i=p_j$ for $i\ne j$. Indeed,
these cases give contributions of order $O(1/L)$ within the summations in \Ep{mn10b}.
Canceling $(a+1)!$ and writing $k=p_{a+1}$ and  $n! = n(n-1)$ we get
\beqa
&&M_{mn}^{R,1}(t) =  mn \int d\Gam\ \sum_{a=0}^{min(m-1,n-1)}
\nonumber\\
&& \times \frac{(m-1)!}{(m-1-a)!a!}  \frac{(n-1)!}{(n-1-a)!} \nonumber\\
&&\times
 \,  [f_1^* q_1^*]^{m-1-a} [f_1 q_1]^{n-1-a}  \EQN{mn10c}\\
&&\times
\sum_{p_1\cdots p_{a}}  |f_{p_1}|^2 \cdots | f_{p_a}|^2
 \nonumber\\
&&\times q^*_{p_1} q_{p_1}\cdots q^*_{p_a} q_{p_a} \sum_k h^*_{k}  f_{k} q^*_k q_k
 \rho(\Gam, 0). \nonumber
\eeqa
Applying the reverse steps from Eqs.  (\ref{mn10b}) to (\ref{mn10}) we get
\beqa
M_{mn}^{R,1}(t)
&=& mn \int d\Gam\,  [q_1^*(t)]^{m-1} [q_1(t)]^{n-1}  \EQN{mn12}  \\
&\times&  \sum_k h^*_k(t) f_{1k} (t)\, q_k^* q_k\,  \rho(\Gam, 0) \nonumber\\
&=& mn \int d\Gam\,  [q_1^*(t)]^{m-1} [q_1(t)]^{n-1}  \nonumber  \\
&\times&  \sum_k h^*_k(t) f_{1k} (t)\, \bra q_k^* q_k \ket \,  \rho(\Gam, 0) + O(1/L) \nonumber
\eeqa
where due to the factorization property \Ep{nfok2} we can take independently the average
\beqa
  \bra q^*_kq_k \ket  = \bra J_k \ket  = \int d\Gam q^*_kq_k \rho(\Gam, 0).
\eeqa
[a similar argument is given in Appendix {\ref{apx:fac}]. Bringing the time dependence back to $\rho$ we get then
\beqa
M_{mn}^{R,1}(t)
&=& mn  \int d\Gam\,  [q_1^*]^{m-1} [q_1]^{n-1} \nonumber \\
&\times&  \sum_k h^*_k(t) f_{1k} (t)\,  \bra J_k \ket \,
\rho(\Gam, t).
 \eeqa %%
For the second term in \Ep{Mmn6} we have $M_{mn}^{R,2}(t) =
[M_{nm}^{R,1}(t)]^*$. Putting everything together in \Ep{nfok1}
with integration by parts and a few straightforward manipulations,
we obtain
\beqa
& &\int d\Gam G(\Gam_1) \frac{\partial}{\partial t} \rho(\Gam,t) \nonumber\\
&=& \int d\Gam  G(\Gam_1)\left\{ iz_1(t) \frac{\partial}{\partial q_1}q_1
-iz_1^*(t) \frac{\partial}{\partial q^*_1} q_1^* \right.\nonumber \\
&+& \left.D(t) \frac{\partial^2}{\partial q_1 \partial q^*_1}  \right\} \rho(\Gam,t).
\EQN{nfok3}
\eeqa
where
\beqa D(t) = \sum_k \left[ h_k^*(t) f_{1k}(t) + {\rm c.c.} \right]
\bra J_k \ket.
 \EQN{Dtres} \eeqa %%
Since $G(\Gam_1)$ is an arbitrary test function, we come to  the  non-Markovian equation (\ref{nfok4}) for the reduced distribution function $\rho_1$. Replacing the explicit form of $h_k^*(t)$ we obtain as well \Ep{Dtres2}.

%%%%%%%%%%%%%%%%%%%
% Appendix
%%%%%%%%%%%%%%%%%%%
\section{Markovian approximation}
\label{apx:mark}

Phenomenological equations  may be obtained from dynamics using  the  so-called Markovian approximation (or the $\lam^2 t$ approximation studied by Van Hove and others \cite{Resibois,VH,BIP,Accardi}), where memory effects are neglected (see also \cite{Leb1,Leb2,Leb3}). This approximation is valid if we take the continuous spectrum limit $L\to\infty$ (with Poincar\'e resonances),  for weak coupling and for time scales where the relaxation process dominates over dressing processes (the particle-bath correlations are negligible \cite{IP62,Gardiner}).

In this appendix we consider the Markovian approximation of the dynamical evolution of the bare particle modes.
We will write the weak coupling approximation  of the coefficients $f_{11}(t)$ and $f_{1k}(t)$, which will give the Markovian approximation of Eqs. (\ref{stand4}) and  (\ref{nfok4}).

We start with \Ep{1-sec-ori}, for $t>0$.  We approximate for weak coupling $\lam\ll 1$,
\beqa \frac{1}{\eta^+ (\ome_k)} \approx \frac{1}{\ome_k -z_1}.
\eeqa %%

We then separate the pole contribution at $\ome_k =z_1$ and the branch cut contribution from each term. The pole contribution gives the exponential decaying part and the cut contribution gives classical Zeno effect and non-exponential behavior \cite{rdamping}. Here we will only consider the pole contributions in \Ep{1-sec-ori}, which amounts to the Markovian approximation. Taking the pole  contribution at $\ome_k = z_1$ in the last two terms of \Ep{1-sec-ori}, we get
\beqa
& &q_1(t) \approx q_1(0) e^{-iz_1 t} + \sum_k \frac{\lam V_k}{\ome_k-z_1} q_{k}(0) e^{-i\ome_k t} \nonumber \\
& &-\sum_k \frac{\lam V_k}{\ome_k -z_1}q_{k}(0) e^{-iz_1 t}. \EQN{q1ap-ori}
\eeqa
Therefore we have
\beqa
& & f_{11}(t) \approx  e^{-iz_1 t},  \nonumber \\
& & f_{1k}(t) \approx    \sum_k \frac{\lam V_k}{\ome_k-z_1}
[e^{-i\ome_k t} - e^{-iz_1t}].
 \EQN{f11m} \eeqa %%
This leads to
\beqa
& &z_1(t) \approx  z_1 , \nonumber \\
& &R(t) \approx   -i  \sum_k \lam V_k e^{-i\ome_k t} q_{k}(0).
\EQN{f12m} \eeqa %%
For $t<0$ we can repeat the same procedure, except that now we choose the $+i\eps$ branch in \Ep{qtk-sec8}.  Then we obtain the following approximate equation
\begin{eqnarray}
 {\dot q}_1(t) \approx -i  z_1 q_1(t) + R(t), \quad t>0 \nonumber\\
 {\dot q}_1(t) \approx  - i z_1^* q_1(t) + R(t), \quad t<0
 \EQN{stand7}
\end{eqnarray}
which has a Langevin form. This equation is expressed in two branches, one for $t>0$ and the other for $t<0$. The change
$-iz_1\Rightarrow -iz_1^*$ corresponds to the change $\gamma\Rightarrow -\gamma$, since $z_1=\omet_1 -i\gamma$. Taken separately, each branch breaks time reversal invariance, while as a whole time reversal invariance is kept [note that in the integrable case we have $\gam=0$  and there is no splitting into two branches].
Comparing \Ep{stand7} with \Ep{a1-sec-lang} we identify $\hz_1=z_1$.
Furthermore, the term $R(t)$ behaves as the white-noise source $\hR(t)$ in the sense that in the pole approximation,
the auto-correlation functions of the variables $\hq_1$ and $q_1$ coincide.  Indeed, the ``noise'' term $R(t)$ in \Ep{f12m} has the same Gaussian property (\ref{R2-sec-lang}) as the noise $\hR(t)$, provided we replace noise averages with averages with ensembles of the form
(\ref{nfok2}) (see discussion below \Ep{mn10a}). Then,   defining
\beqa
&& q_{1a}(t) \equiv q_1(0) e^{-iz_1 t}, \EQN{mcorr}\\
&&  q_{1r}(t) \equiv e^{-iz_1 t} \int_0^t dt'R(t') e^{iz_1 t'} \nonumber\\
&& =   \sum_k \frac{\lam V_k}{\ome_k-z_1}  [e^{-i\ome_k t} - e^{-iz_1t}] q_k(0)
\nonumber\\
\eeqa
and following the same steps as in Eqs. (\ref{a1t-sec-lang})- (\ref{arr-sec-lang}) we get
\beqa \bra q_{1r}^{*k}(t) q_{1r}^l(t) \ket = l! \delta_{kl} \bra
q_{1r}^*(t)q_{1r}(t) \ket^l.
 \EQN{mcorr2} \eeqa %%
In general we have (with $q_{k0} \equiv q_k(0)$)
\beqa
& &\bra q_{1r}^* (t+\tau) q_{1r}(t) \ket \nonumber \\
& & \approx \bra \sum_k \frac{\lam V_k}{\ome_k-z_1^*} q_{k0}^* (e^{i\ome_k (t+\tau)}  - e^{iz_1^* (t+\tau)}) \right. \nonumber \\
& &\left. \times \sum_{l} \frac{\lam V_{l}}{\ome_{l}-z_1} q_{l0}
(e^{-i\ome_{l}t}  - e^{-iz_1 t} ) \ket.
 \EQN{4-sec-ori}
\eeqa
 For the normalized thermal field ensemble we  have
\beqa
\bra q_{k0}^* q_{k'0} \ket = \delta_{kk'} \frac{k_B T}{\ome_k}.
\EQN{5-sec-ori} \eeqa
Using this result and going to the continuous limit we obtain from
\Ep{4-sec-ori}
\beqa
& & \bra q_{1r}^* (t+\tau) q_{1r}(t) \ket \nonumber \\
& &\approx \int_0^\infty dw \frac{\lam^2 v_w^2}{|\ome-z_1|^2} \frac{k_B T}{\ome} \nonumber \\
& &\times (e^{i\ome \tau} - e^{-iz_1 t}e^{i\ome (t+\tau)}- e^{iz_1^* (t+\tau)}
e^{-i\ome t} +e^{iz_1^* \tau} e^{-2\gam t}). \nonumber \\
\EQN{6-sec-ori} \eeqa
For $\gamma\ll\omet_1$ the integrand is sharply peaked around
$\ome=\omet_1$.  We separate the  pole  and the cut contributions
to the integral,  rewriting \Ep{6-sec-ori} as
\beqa
& & \bra q_{1r}^* (t+\tau) q_{1r}(t) \ket \EQN{7-sec-ori'}  \\
& &\approx \bra q_{1r}^* (t+\tau) q_{1r}(t) \ket_{pole} + \bra
q_{1r}^* (t+\tau) q_{1r}(t) \ket_{cut}\nonumber \eeqa
where
\beqa
& & \bra q_{1r}^* (t+\tau) q_{1r}(t) \ket_{pole} \nonumber \\
& &=  \int_{-\infty}^{\infty} d\ome \frac{\lam^2 v_{\ome}^2 }{|\ome-z_1|^2} \frac{k_B T}{\ome} \nonumber \\
& &\times (e^{i\ome \tau} -  e^{-iz_1 t}e^{i\ome (t+\tau)}-
e^{iz_1^* (t+\tau)} e^{-i\ome t} + e^{iz_1^* \tau} e^{-2\gam t}),
\nonumber\\
\EQN{7-sec-oripole} \eeqa
\beqa
& & \bra q_{1r}^* (t+\tau) q_{1r}(t) \ket_{cut} \nonumber \\
& &= -\int_{-\infty}^0 d\ome \frac{\lam^2 v_\ome^2}{|\ome-z_1|^2} \frac{k_B T}{\ome} \nonumber \\
& &\times (e^{i\ome \tau} - e^{-iz_1 t}e^{i\ome (t+\tau)}-
e^{iz_1^* (t+\tau)}e^{-i\ome t} +e^{iz_1^* \tau} e^{-2\gam t}).
\nonumber\\
\EQN{7-sec-oricut} \eeqa
Using
\beqa
v_{z_1} &\approx& v_{\omet_1}, \quad \quad \frac{k_B T}{z_1} \approx \frac{k_B T}{\omet_1},\nonumber\\
  \gam &\approx& \pi \lam^2 v_{\omet_1}^2,
 \EQN{vapprox}
\eeqa
the pole contributions at $\ome = z_1,\, \ome = z_1^*$ give
\beqa \bra q_{1r}^* (t+\tau) q_{1r}(t) \ket_{pole} \approx
\frac{k_B T}{\omet_1} e^{iz_1^*\tau} (1-e^{-2\gam t}).
\EQN{q1cor-ori} \eeqa
 The cut part  involves the tail of the Lorentzian
distribution $1/|\ome-z_1|^2={1}/((\ome-\omet_1)^2 + \gam^2)$. It
gives
\beqa
& & \bra q_{1r}^* (t+\tau) q_{1r}(t) \ket_{cut} \nonumber \\
& &\approx \int_{0}^\infty d\ome \frac{\lam^2 v_{-\ome}^2}{|\ome+z_1|^2} \frac{k_B T}{\ome} \nonumber \\
& &\times (e^{-i\ome \tau} - e^{-iz_1 t}e^{-i\ome (t+\tau)}- e^{iz_1^* (t+\tau)}e^{i\ome t} +e^{iz_1^* \tau} e^{-2\gam t}). \nonumber \\
\EQN{8-sec-ori} \eeqa
 In weak coupling
case ($\gam \ll \omet_1$),  the cut contribution is much smaller
than the pole contribution. Replacing the result (\ref{q1cor-ori}) in \Ep{mcorr2} with $\tau=0$ we obtain the same correlation \Ep{avg-sec-lang} obtained from the phenomenological Langevin equation.

We have also
\beqa D(t) \approx \frac{2\gamma k_B T}{\omet_1}.
 \eeqa %%
Hence we recover the Markovian Fokker-Planck equation (\ref{int-fp-fac3}) with $\hz_1 = z_1$.

%%%%%%%%%%%%%%%%%%%%%%
% AppendixA
%%%%%%%%%%%%%%%%%%%%%%
\section{Proof of \Ep{dsing}}
\label{app:dsing}
In this appendix we show, by perturbation expansion that \Ep{dsing} is satisfied for the equilibrium Gibbs ensemble. The same relation
is valid for the class of   ensembles (generally non-equilibrium ensembles) having delta-singularities in the wave numbers, of which the Gibbs ensemble is a special case \cite{IP62,PP96,PP97}. The Gibbs ensemble is given by
\beqa
 \rho^{\rm eq} = C  \exp (-\beta H)
 \EQN{rsing1}
\eeqa
where $C$ is a normalization constant, such that
\beqa
 \int d\Gamma \rho^{\rm eq}(\Gamma) = 1.
 \EQN{rsing2}
\eeqa
In the perturbation expansion we have
\beqa
 \rho^{\rm eq} = C  \exp (-\beta H_0)\left(1+\lam V + \frac{1}{2!} \lam^2V^2 + \cdots
 \right).
 \EQN{rsing4}
\eeqa
Due to the angle integrations, only diagonal monomials are nonzero
\beqa
 \int d\Gamma \prod_r q_r^{*m_r} q_r^{n_r}  \exp (-\beta H_0) \propto \prod_r
 \del_{m_r,n_r}.
 \EQN{rsing5}
\eeqa
Considering the explicit form of the potential $V$ we then have
\beqa
& &\bra J_k \ket =  \int d\Gamma J_k  \rho^{\rm eq}  \nonumber\\
& &= \frac{\int d\Gamma J_k \exp(-\beta H_0)}{\int d\Gamma \exp(-\beta H_0)} + O(\lam^2) \nonumber\\
& &= \frac{\int_0^\infty dJ_k J_k \exp(-\beta \ome_k J_k )}{\int_0^\infty dJ_k  \exp(-\beta \ome_k J_k )} + O(\lam^2) \nonumber\\
& & = \frac{1}{\beta \ome_k} + O(\lam^2),
 \EQN{rsing6}
\eeqa
\beqa
& &\bra q_k^* q_{k'} \ket =  \frac{\int d\Gamma q_k^* q_{k'} (\lam^2 V^2/2) \exp(-\beta H_0)}{\int d\Gamma \exp(-\beta H_0)} + O(\lam^4)  \nonumber\\
& & = \lam^2 V_k V_{k'}   \frac{1}{\beta \ome_1}  \frac{1}{\beta
\ome_k}  \frac{1}{\beta \ome_{k'}}+ O(\lam^4).
 \EQN{rsing7}
\eeqa
Noting that $V_k \sim L^{-1/2}$, and $\sum_{k'} \sim O(L)$, we obtain the result (\ref{dsing}). One can check as well that terms of higher order in $\lam V$ retain the same volume dependences.

For an ensemble of the form \Ep{nfok2}, we  have
$\bra q_k^* q_{k'} \ket = 0$, while $\bra J_k \ket \sim O(1)$.  This is a special case of \Ep{dsing}.

%%%%%%%%%%%%%%%%%%%%%%
% Appendix B
%%%%%%%%%%%%%%%%%%%%%%
\section{$\Lam$ and preservation of the measure}
\label{app:r}

From the requirement (2) in the Introduction, $\Lam$ preserves the measure of phase space. This means that
\beqa
 & & \int d\Gamma \Lam \rho = 1, \EQN{meas3'}\\
& & \int d\Gamma \Lam^\dagger \rho = 1
 \EQN{meas3}
\eeqa
for any normalized ensemble $\rho$.
The first equality is easily shown, since it may be written as
\beqa
  \int d\Gamma (\Lamd 1)  \rho = 1
 \EQN{meas30}
\eeqa
where $1$ is the unit operator. Since  $\Lamd$  can be expressed as a perturbation expansion,
\beqa
  \Lamd  = 1 + O(\lam L_V)
 \EQN{meas31}
\eeqa
and $L_V 1 = 0$ we conclude that $\Lamd 1 = 1$, from which \Ep{meas3'} follows.  Similarly, one can show the second equality (\ref{meas3}).

So, \Ep{meas3}  should be satisfied for the specific $\Lam$ we have constructed in Sec. \ref{sec:Lambda}.
This condition  will allow us to derive the relation (\ref{rdef}). Consider  the ensemble
\beqa
\rho = C_1 q_1^* q_1 \exp(-J/J_0)  \EQN{meas1}
\eeqa
where $C_1$ is the normalization factor given by
\beqa
 C_s = \Big[\int d\Gamma  q_s^* q_s \exp(-J/J_0)\Big]^{-1}  \EQN{meas1'}
\eeqa
with
\beqa
 J = \sum_{s} q_s^* q_s
 \EQN{meas2}
\eeqa
and $J_0$ a constant that makes the argument of the exponential dimensionless. The factor $\exp(-J/J_0)$ ensures the existence of a finite norm of $\rho$, (see the Segal-Bargmann representation in  \cite{Klauder,rdamping}). The total action  $J$ is an invariant of motion, because we have $L_0 J =0$ and $L_V J =0$.    Using \Ep{meas31}, we get
\beqa \Lamd J = J, \quad \Lamd \exp(-J/J_0) = \exp(-J/J_0).
 \EQN{meas4}
\eeqa
The operator $L_V$ is a differential operator. Applying the chain rule of differentiation and \Ep{meas4} we conclude that
\beqa \Lamd q_1^* q_1 \exp(-J/J_0) =  (\Lamd q_1^* q_1)
\exp(-J/J_0).
 \EQN{meas5}
\eeqa
Inserting the ensemble \Ep{meas1} in \Ep{meas3}  and using \Ep{Lam-J1} we get
\beqa
& & C_1 \int d\Gamma |N_1|[q_1^*q_1 + \lam^2 \sum_k (rc_k^2 + {\rm c.c.}) q_k^*q_k] \nonumber \\
& &\times \exp(-J/J_0) = 1.
 \EQN{meas6}
\eeqa
where the off-diagonal terms such as $q_1^* q_k$ appearing in the product $\Qt_1^*\Qt_1$ in \Ep{Lam-J1} vanish due to the integration over  angles in  phase space. We can write \Ep{meas6} as
\beqa C_1   |N_1|[ C_1^{-1} + \lam^2\sum_k (rc_k^2 + {\rm c.c.})
C_k^{-1}]  =  1.
 \EQN{meas6'}
\eeqa
 Since
$C_1=C_k$ for any $k$ (see \Ep{meas1'}), \Ep{meas6}  leads to
\beqa
 |N_1|[1 + \lam^2 \sum_k  (rc_k^2 + {\rm c.c.})] = 1.
  \EQN{meas7}
\eeqa
This equation plus the condition $r+r^*=1$ yield the result (\ref{rdef}).  With this result we can write
\beqa
\Lamd q_1^* q_1 = Q^{(0)} \Qt_1^* \Qt_1 + P^{(0)} (r \Qt_1^* Q_1 + \rm{c.c.})
  \EQN{LamGam}
\eeqa
where $P^{(0)}$ is the projector to angle-independent monomials (such as $q_s^*q_s$) and $ Q^{(0)} =1-P^{(0)}$. This shows that the transformed product of particle modes can be simply written as a combination of  renormalized particle modes (i.e., Gamow modes).

The derivation followed here is similar to the derivation followed in Ref. \cite{OPP}, where
 we used the $\Lam$ transformation to define dressed unstable states in quantum mechanics. The only difference is that in \cite{OPP} the relation $r+r^*=1$ was derived from the requirement that the dressed unstable state has an energy fluctuation of the order of the inverse lifetime. This fluctuation is a quantum effect. Here we are dealing with classical mechanics, so we postulate $r+r^*=1$ as a basic condition. An alternative derivation, presented in Appendix A of Ref. \cite{POP2001}, started with the analogue of \Ep{LamGam}, as a postulate. All the derivations give the same result (\ref{rdef}). Note that a condition different from $r+r^* = 1$ would not allow us to express  $\Lamd q_1^* q_1$ as a combination of Gamow modes only, and it would lead to energy fluctuations different from the inverse lifetime in the quantum case, which would be unphysical.

The preservation of the measure can be proven for more general ensembles involving monomials of particle modes, which we have considered in Sec. \ref{sec:Lambda}.  Indeed, from the relation $\sum_k c_k^*c_k = -1$ (see \cite{OPP}) we have (see Eqs. (\ref{meas7}), (\ref{bkdef}))
\beqa \sum_k b_k = 1.
  \EQN{meas8}
\eeqa
Using this relation as well as the expression (\ref{LAM-MAIN}) one can show that for $\rho = C_{mn} q_1^{*m} q_1^n \exp(-J/J_0)$
with $C_{mn}$ a normalization constant, we have
\beqa
 \int d\Gamma \Lam^\dagger \rho =  \int d\Gamma  \rho = \del_{mn}.
  \EQN{meas9}
\eeqa
The proof (which we will omit here) uses the relation $ \int d\Gamma \Qt_1^{*m} \Qt_1^n \exp(-J/J_0) = 0$. This follows from the fact that both $\Qt_1^{*m} \Qt_1^n$ and $\exp(-J/J_0)$ are eigenfunctions of $L_H$ with different eigenvalues, which implies their orthogonality.

%%%%%%%%%%%%%%%%%%%%%%
% Appendix C
%%%%%%%%%%%%%%%%%%%%%%
\section{Proof of analyticity of $\Lam$}
\label{apx:analytic}
In this appendix we show that \Ep{LAM-MAIN} removes all the non-analytic $|c_k|^2$ terms, replacing them   by $\xi_k = r c_k^2 +c.c$. First we derive recursive formulas to calculate $\Lamd q^{*m}_1q^n_1$. We start with \Ep{LAM-MAIN} for $m \geq n$
\beqa
\Lam^{\dagger}q^{*m}_1q^n_1  = \sum_{l=0}^n \frac{m!n!}{(m-l)!(n-l)!l!}\Qt^{*m-l}_1 \Qt^{n-l}_1Y^l.  \EQN{recursive1}
\eeqa
(the $n>m$ case can be calculated by taking the complex conjugate of \Ep{recursive1}). We have as well
\beqa
& &\Lam^{\dagger}q^{*m+1}_1q^n_1  \nonumber \\
& &= \sum_{l=0}^n
 \frac{(m+1)!n!}{(m+1-l)! (n-l)!l!} \Qt^{*m+1-l}_1 \Qt^{n-l}_1Y^l.
 \EQN{recursive3}
\eeqa
For $l>0$ we have the identity
\beqa
 \frac{(m+1)!}{(m+1-l)! l!}
 =
 \frac{m!}{(m-l)! l!} + \frac{m!}{(m+1-l)!(l-1)!}.
 \EQN{recursive4}
\eeqa
Inserting this in \Ep{recursive3} we get
\beqa
& &\Lam^{\dagger}q^{*m+1}_1q^n_1  = \Qt^{*m+1}_1 \Qt^{n}_1 \nonumber\\
& & + \sum_{l=1}^n
 \Big[\frac{m!}{(m-l)! l!} + \frac{m!}{(m+1-l)!(l-1)!}\Big] \nonumber \\
& &\times\frac{n!}{(n-l)!}
 \Qt^{*m+1-l}_1 \Qt^{n-l}_1Y^l.
 \EQN{recursive5}
\eeqa
The first term plus the second term give
\beqa
 (\Lam^{\dagger}q^{*m}_1q^n_1) \Lam^{\dagger} q^*_1
 \EQN{recursive6}
\eeqa
(note that $ \Lam^{\dagger} q^*_1 = \Qt^{*}_1$).
The third term may be written as (with $l'=l-1$)
\beqa
& &\sum_{l'=0}^{n-1}
  \frac{m!}{(m-l')!(l')!} \frac{n(n-1)!}{(n-l'-1)!}
 \Qt^{*m-l'}_1 \Qt^{n-l'-1}_1Y^{l'+1} \nonumber \\
& &= nY \Lam^\dagger q^{*m}_1q^{n-1}_1.
 \EQN{recursive7}
\eeqa
Therefore
\beqa
& &\Lam^{\dagger}q^{*m+1}_1q^n_1  \EQN{recursive8} \\
& & = (\Lam^{\dagger}q^{*m}_1q^n_1) \Lam^{\dagger} q^*_1
 + nY \Lam^\dagger q^{*m}_1q^{n-1}_1 \,\,\,\,\,(m\ge n).
 \nonumber
\eeqa
For $m>n$ we have, from \Ep{recursive1},
\beqa
\Lam^{\dagger}q^{*m}_1q^{n+1}_1  &=& \sum_{l=0}^n
\frac{m!(n+1)! }{(m-l)!(n+1-l)!l!} \nonumber\\
&\times& \Qt^{*m-l}_1 \Qt^{n+1-l}_1 Y^l.
 \EQN{recursive9}
\eeqa
Using \Ep{recursive4} we get
\beqa
& & \Lam^{\dagger}q^{*m}_1q^{n+1}_1 \nonumber \\
& &= \Qt^{*m}_1 \Qt^{n+1}_1 + \frac{m!}{(m-n-1)!}\Qt^{*m-n-1}_1 Y^{n+1}
 \nonumber\\
 & &+ \sum_{l=1}^n  \frac{m!}{(m-l)!}
 \Big[\frac{n!}{(n-l)! l!} + \frac{n!}{(n+1-l)!(l-1)!}\Big] \nonumber \\
& &\times \Qt^{*m-l}_1 \Qt^{n+1-l}_1Y^l.
 \EQN{recursive10}
\eeqa
Adding the first and the third terms we get
\beqa
  (\Lam^{\dagger} q^{*m}_1 q^{n}_1)  \Lam^{\dagger} q_1
 \EQN{recursive11}
\eeqa
(note that $ \Lam^{\dagger} q_1 = \Qt_1$).
Adding the second and fourth terms we get (with $l'=l-1$)
\beqa
& & \sum_{l'=0}^n \frac{m(m-1)!}{(m-l'-1)!}
 \frac{n!}{(n-l')!(l')!} \Qt^{*m-l'-1}_1 \Qt^{n-l'}_1Y^{l'+1} \nonumber \\
& & =
 mY \Lam^\dagger q^{*m-1}_1q^{n}_1.
 \EQN{recursive12}
\eeqa
Therefore
\beqa
& &\Lam^{\dagger}q^{*m}_1q^{n+1}_1 \EQN{recursive13} \\
& & = (\Lam^{\dagger}q^{*m}_1q^n_1) \Lam^{\dagger} q_1
 + mY \Lam^\dagger q^{*m-1}_1q^{n}_1  \,\,\,\,\,(m>n). \nonumber
\eeqa
Eqs. (\ref{recursive8}) and (\ref{recursive13}) plus their complex conjugates
permit one to construct $\Lam^{\dagger}q^{*m}_1q^{n}_1$ recursively.

Now we prove the analyticity of $\Lamd q_1^{*m} q_1^n$ at $\lam=0$ from the recursive relations.  In the recursive relation, we show that if the lower order terms in $m$ and $n$ like $\Lamd q^{*m}q_1^n$,  $\Lamd q_1^{*m}q_1^{n-1}$ and   $\Lamd q_1^{*m-1}q_1^{n-1}$
  are analytic, then the higher order terms $\Lamd q^{*m+1}q_1^n$ and $ \Lamd q_1^{*m}q_1^{n}$ are also analytic. Then from mathematical induction,
the analyticity of $\Lamd q^{*m}_1 q_1^n$ is proved for general $m$ and $n$ (the $m <n$ case can be shown in the same way).
 In \Ep{recursive8},
\beqa
& &\Lamd q_1^{*m+1}q_1^n = (\Lamd q^{*m}q_1^n)\Lamd q_1^* + n Y \Lamd q_1^{*m}q_1^{n-1} \nonumber \\
& &= \bigg(\sum_{l=0}^n \frac{m!n!}{(m-l)!(n-l)!l!} \Qt_1^{*m-l}\Qt_1^{n-l}Y^l\Bigg) \Qt_1^* \nonumber \\
& &+ n Y \Bigg( \sum_{l=0}^{n-1} \frac{m!(n-1)!}{(m-l)!(n-1-l)!l!} \Qt_1^{*m-l}\Qt_1^{n-1-l} Y^l\Bigg) \nonumber \\
 \EQN{2-anal-apx}
\eeqa
 Suppose that the quantities inside large parenthesis are analytic in $\lam$. The additional non-analytic terms appear whenever additional products $\Qt^*_1 \Qt_1$ appear. Since
\beqa
\Qt_1 = N_1^{1/2}(q_1 + \lam \sum_k c_k q_k),
\eeqa
each $\Qt^*_1 \Qt_1$ produces a $|c_k|^2$ term, which is non-analytic in $\lam$. Let us denote the non-analytic part of a function $f(\lam)$ as $Fn(f(\lam))$.
The non-analytic part in the first term in the right hand side of \Ep{2-anal-apx} is made by the additional $\Qt^*_1$ multiplied by $\Qt_1^{n-l}$, which generates $n-l$ terms $|c_k|^2$:
\beqa
& &Fn\Bigg[ (\sum_{l=0}^n \frac{m!n!}{(m-l)!(n-l)!l!} \Qt^{*m-l}_1\Qt_1^{n-l} Y^l) \Qt^*_1 \Bigg] \nonumber \\
& &= \sum_{l=0}^n \frac{m!n!}{(m-l)!(n-l)!l!}\Qt_1^{*m-l}
\Qt_1^{n-l-1}Y^l \nonumber \\
& &\times (n-l)\lam^2 |N_1| \sum_k |c_k|^2 q_k^* q_k \nonumber \\
& & = \sum_{l=0}^{n-1}\frac{m!n!}{(m-l)!(n-l-1)!l!}\Qt_1^{*m-l} \Qt_1^{n-l-1}Y^l \nonumber \\
& &\times \lam^2 |N_1| \sum_k |c_k|^2 q_k^* q_k.
\eeqa
The non-analytic part in the second term in the right hand side of \Ep{2-anal-apx} is coming from $Y$. Since
\beqa
& &Y=\sum_k b_k q^*_k q_k \nonumber \\
& &= \sum_k \lam^2 |N_1|(-|c_k|^2 + rc_k^2 + r^* c_k^{*2}) q_k^* q_k,
\eeqa
 the non-analytic function $|c_k|^2$ appears inside $Y$.
\beqa
& &Fn\Bigg[ nY\sum_{l=0}^{n-1} \frac{m!(n-1)!}{(m-l)!(n-1-l)!l!}\Qt^{*m-l}_1\Qt_1^{n-1-l}Y^l \Bigg ] \nonumber \\
& &= -\lam^2 |N_1| \sum_k |c_k|^2 q^*_k q_k \nonumber \\
& &\times \sum_{l=0}^{n-1}\frac{m!n!}{(m-l)!(n-1-l)!a!}\Qt_1^{*m-l}\Qt_1^{*n-1-l} Y^l.
\eeqa
 The non-analytic parts from the first term and second term in \Ep{2-anal-apx} exactly cancels out. So, the left hand side of \Ep{2-anal-apx} is analytic in $\lam$.  Note that terms of the form $|c_k|^{2n}$ with $n>1$ give $O(1/L)$ contributions and thus they are negligible.

 Next, we show that the left hand side of \Ep{recursive13} is analytic in $\lam$. The non-analytic part of the first term in the right hand side of \Ep{recursive13} is
\beqa
& &Fn\Bigg[ (\Lamd q_1^{*m}q_1^{n}) \Lamd q_1 \Bigg ] \nonumber \\
& &= Fn\Bigg [ (\sum_{l=0}^{n} \frac{m!n!}{(m-l)!(n-l)!l!} \Qt^{*m-l}_1\Qt_1^{n-l} Y^l) \Qt_1 \Bigg] \nonumber \\
& & =  \sum_{l=0}^{n} \frac{m!n!}{(m-l)!(n-l)!l!} \Qt^{*m-l-1}_1\Qt_1^{n-l} Y^l \nonumber \\
& &\times (m-l) \lam^2 |N_1| \sum_k |c_k|^2 q^*_k q_k \nonumber \\
& &=  \sum_{l=0}^{n} \frac{m!n!}{(m-l-1)!(n-l)!l!} \Qt^{*m-l-1}_1\Qt_1^{n-l} Y^l \nonumber \\
& &\times \lam^2 |N_1| \sum_k |c_k|^2 q^*_k q_k
\eeqa
The non-analytic part of the second term in the right hand side of \Ep{recursive13} is
\beqa
& &Fn\Bigg[ m Y \Lamd q_1^{*m-1}q_1^{n} \Bigg] \\
& &= -m \lam^2 |N_1|\sum_k |c_k|^2 q^*_k q_k \nonumber \\
& &\times \sum_{l=0}^{n} \frac{(m-1)!n!}{(m-l-1)!(n-l)!l!} \Qt_1^{*m-1-l}\Qt_1^{n-l} Y^l \nonumber \\
& & =  - \lam^2 |N_1|\sum_k |c_k|^2 q^*_k q_k \nonumber \\
& &\times \sum_{l=0}^{n} \frac{m!n!}{(m-l-1)!(n-l)!l!} \Qt_1^{*m-1-l}\Qt_1^{n-l} Y^l. \nonumber
\eeqa
Again, the non-analytic parts of the first and second terms of \Ep{recursive13} exactly cancel out. The right hand side of \Ep{recursive13} is analytic in $\lam$.
 Therefore from the mathematical induction $\Lamd q_1^{*m} q_1^n$ is analytic in $\lam$.

%%%%%%%%%%%%%%%%%%%%%%
% Appendix E
%%%%%%%%%%%%%%%%%%%%%%
\section{ Calculation of the noise constant $\hR_{\c}$}
\label{apx:AcBc}
 In this appendix we determine the noise constant $\hR_c$. We assume that the noise $\hR(t)$  comes from the thermal bath with temperature $T$. In this case, we expect that the system reaches thermal equilibrium for $t \rightarrow \infty$. Furthermore, from the equipartition theorem we expect that
\beqa
\frac{1}{2} {\hat m} \home_1^2 \bra \hx_1^2 \ket_{eq} = \frac{\bra \hp_1^2 \ket_{eq} }{2 {\hat m}} = \frac{1}{2} k_B T, \EQN{xp-sec-lang}
\eeqa
where $k_B$ is Boltzmann's constant. Substituting the relations
\beqa
& &\hx_1(t) = \sqrt{\frac{1}{2m_p\home_1}} (\hq_1(t)+\hq_1^*(t)), \nonumber \\
& & \hp_1(t) = -i\sqrt{\frac{m_p\home_1}{2}} (\hq_1(t)-\hq_1^*(t))
\eeqa
into \Ep{xp-sec-lang}, we get the conditions
\beqa
& &\bra \hq_1^2(t) \ket_{eq} + \bra \hq_1^{*2}(t) \ket_{eq} = 0, \EQN{con-sec-0}\\
& & \home_1 \bra \hq_1^*(t) \hq_1(t) \ket_{eq}= k_B T. \EQN{con-sec-lang}
\eeqa
On the other hand we have
\beqa
& &\bra \hq^*_1(t) \hq_1(t) \ket_{eq} \nonumber \\
& &= \lim_{t\rightarrow \infty} (\bra \hq^*_{1a}(t) \hq_{1a}(t) \ket + \bra \hq^*_{1r}(t) \hq_{1r}(t) \ket) \nonumber \\
& &= \lim_{t\rightarrow \infty} \bra  \hq^*_{1r}(t) \hq_{1r}(t) \ket,
\eeqa
\beqa
& &\bra \hq^*_{1a}(t) \hq_{1r}(t) \ket \nonumber \\
& &= \bra e^{-2\hgam t} \int_0^t \int_0^t dt_1 dt_2 \hR(t_1) \hR^*(t_2) \right. \nonumber \\
& &\left. \times e^{i\hz_1t_1-1\hz_1^* t_2)} \ket \nonumber \\
& & = e^{-2\hgam t} \int_0^t dt_1 \hR_c^2 e^{2\hgam t_1}  \frac{\hR_c^2(1-e^{-2\hgam t})}{2\hgam}.
\eeqa
Substituting this to  \Ep{con-sec-lang}, we get
\beqa
\hR_c^2 = \frac{2\hgam k_B T}{\home_1}.
\eeqa
%%
%%%%%%%%%%%%%%%%%%%%%%
% Appendix F
%%%%%%%%%%%%%%%%%%%%%%
\section{Calculation of the moments}
\label{apx:moments}
In this appendix we calculate the moments in \Ep{moments}.
We have
\beqa
& &\int d\Gam (q_1^*-q_1^{'*})^m (q_1-q_1')^n \tht (\Gam) \delta (\Gam-\Gam')  \EQN{1-moment-apx}  \\
& &= \int d\Gam [\tht^{\dagger}(\Gam) (q_1^*-q_1^{'*})^n (q_1-q_1')^m ]^* \delta (\Gam-\Gam') \nonumber \\
& & =-\int d\Gam [(\Lamd)^{-1}L_H \Lamd (q_1^*-q_1^{'*})^m (q_1-q_1')^n ] \delta (\Gam-\Gam'), \nonumber
 \eeqa
where we used the relation $L_H^{\dagger} = L_H$ and $L_H^*=
-L_H$. The quantity inside the brackets in \Ep{1-moment-apx}
[which we call $I$] is %%
\beqa
& &I =(\Lamd)^{-1}L_H \Lamd (q_1^*-q_1^{'*})^m (q_1-q_1')^n \nonumber \\
& &= -i\frac{d}{dt}(\Lamd)^{-1} e^{iL_H t} \Lamd (q_1^*-q_1^{'*})^m (q_1-q_1')^n \mid_{t=0}  \nonumber \\
& &= \sum_{l=0}^m \sum_{j=0}^n (-q_1^{'*})^l (-q_1')^j \frac{m!n!}{(m-l)!(n-j)!l!j!} \nonumber \\
& &\times (-i\frac{d}{dt}) (\Lamd)^{-1}e^{iL_H t} \Lamd q_1^{*m-l} q_1^{n-j}\mid_{t=0}
\eeqa
Using \Ep{Lam-sec-lang}, we have
\beqa
& &I = \sum_{l=0}^m \sum_{j=0}^n (-q_1^{'*})^l (-q_1')^j \frac{m!n!}{(m-l)!(n-j)!l!j!} (-i\frac{d}{dt}) \nonumber \\
& &\times \sum_{a=0}^{min(m-l,n-l)} \frac{(m-l)!(n-l)!}{(m-l-a)!(n-j-a)!a!} \nonumber \\
& &\times e^{i( (m-l)z_1^* - (n-j)z_1)t} q_1^{*m-l-a} q_1^{n-j-a} Y^{a} \nonumber \\
& &\times (e^{2\gam t}-1)^a \mid_{t=0}. \EQN{2-moment-apx}
\eeqa
Because of the $(e^{2\gam t}-1)^a$ term, the only non-vanishing terms in \Ep{2-moment-apx} at $t=0$ are for $a=0$ or $a=1$. So the above equation becomes
\beqa
& &I = \sum_{l=0}^m \sum_{j=0}^n (-q_1^{'*})^l (-q_1')^j \frac{m!n!}{l!j!(m-l)!(n-j)!} \nonumber \\
& &\times (-i\frac{d}{dt}) (e^{iz_1^* t}q_1^*)^{m-l} (e^{-iz_1 t}q_1)^{n-j} \mid_{t=0} \nonumber \\
& & + \sum_{l=0}^{m-1} \sum_{j=0}^{n-1}  (-q_1^{'*})^l (-q_1')^j\frac{m!n!}{l!j!(m-l-1)!(n-j-1)!} \nonumber \\
& &\times (-i\frac{d}{dt}) e^{i(z_1-z_1^*)t}  (e^{iz_1^* t}q_1^*)^{m-l-1} (e^{-iz_1 t}q_1)^{n-j-1} Y\nonumber \\
& &\times (e^{2\gam t}-1)\mid_{t=0} \nonumber \\
& &=  (-i\frac{d}{dt}) (e^{iz_1^* t}q_1^* -q_1^{'*})^m (e^{-iz_1 t}q_1 -q_1')^n\mid_{t=0} \nonumber \\
& &+  (-i\frac{d}{dt}) m n Y (1-e^{-2\gam t}) \nonumber \\
& &\times (e^{iz_1^* t} q_1^* -q_1^{'*})^{m-1} (e^{-iz_1 t} q_1 -q_1')^{n-1}\mid_{t=0}.
 \EQN{3-moment-apx}
\eeqa
Substituting \Ep{3-moment-apx} into \Ep{1-moment-apx} and integrating with
$\delta (\Gam-\Gam')$, we get \Ep{moments}.

%%%%%%%%%%%%%%%%%%%%%%
% Appendix G
%%%%%%%%%%%%%%%%%%%%%%
\section{Factorization property  }
\label{apx:fac}
We show the factorization of \Ep{fac-sec-fp} when $\rhot(\Gam, 0)$ has the form
\beqa \rhot(\Gam,0) = g_1(q_1^*,q_1)\prod_k g_k (q_k^*,q_k).
 \eeqa
In \Ep{fac-sec-fp}, by integrating by parts, we can write
\beqa
& &\int d\Gam G(q_1,q^*_1) \frac{\partial^2}{\partial q_1 \partial q^*_1}\sum_k b_k J_k \rhot(\Gam,t) \nonumber \\
& &= \int d\Gam\frac{\partial^2}{\partial q_1 \partial q^*_1}   G(q_1,q^*_1)\sum_k b_k J_k \rhot(\Gam,t) \EQN{189-apx} \\
& &=  \int d\Gam\Bigg(\frac{\partial^2}{\partial q_1 \partial q^*_1}   G(q_1,q^*_1)\Bigg)\sum_k b_k J_k e^{-i\tht t} \rhot(\Gam,0). \nonumber
\eeqa
Let us expand
%5
\beqa
\frac{\partial^2}{\partial q_1 \partial q^*_1}   G(q_1,q^*_1) = \sum_{m,n} G_{mn} q_1^{*m} q_1^n. \EQN{190-apx}
\eeqa
 We have
\beqa
& &\int d\Gam q_1^{*m}q_1^n \sum_k b_k J_k e^{-i\tht t} \rhot (\Gam,0) \nonumber \\
& &= \int d\Gam [(e^{-i\tht t})^{\dagger}q_1^{m}q_1^{*n} \sum_k b_k J_k]^* \rhot (\Gam,0).
\eeqa
Since
\beqa
L_0 \sum_k b_k J_k =0,\quad L_V \sum_k b_k J_k = O(1/\sqrt{L})
\eeqa
and $\Lam$ is expressed in terms of $L_0$ and $L_V$, $\tht = \Lam L_H \Lam^{-1}$ treats $\sum_k b_k J_k$ like constant. Neglecting $O(1/\sqrt{L})$ terms,  we can write
\beqa
& &\int d\Gam q_1^{*m}q_1^n \sum_k b_k J_k e^{-i\tht t} \rhot (\Gam,0) \nonumber \\
& &=
\int d\Gam \sum_k b_k J_k [(e^{-i\tht t})^{\dagger}q_1^{m}q_1^{*n}]^* \rhot (\Gam,0). \EQN{1-apx6}
\eeqa
In \Ep{1-apx6},
$[(e^{-i\tht t})^{\dagger}q_1^{m}q_1^{*n}]^*$ can be written as (see \Ep{Lam-sec-lang})
\beqa
& &[(e^{-i\tht t})^{\dagger}q_1^{m}q_1^{*n}]^* = (\Lamd)^{-1} \bigg(e^{iL_H t} \Lamd q_1^{*m}q_1^n \bigg) \nonumber \\
& &= \sum_{l=0}^{min(m,n)} \frac{m!n!}{(m-l)!(n-l)!l!} \nonumber \\
& &\times e^{i(mz_1^*-nz_1)t} q_1^{*m-l}q_1^{n-l} Y^l (e^{2\gam
t}-1)^l.
 \EQN{2-apx6}
\eeqa
 Since
\beqa
\sum_k b_k J_k (\sum_k b_k J_k)^l= \sum_k b_k J_k (\sum_{k' \ne k} b_{k'} J_{k'})^l +  O(1/L),\nonumber \\
\eeqa
 we can write
\beqa
& & \sum_k b_kJ_k [(e^{-i\tht t})^{\dagger}q_1^{m}q_1^{*n}]^* \nonumber \\
& &= \sum_k b_k J_k  [(e^{-i\tht t})^{\dagger}q_1^{m}q_1^{*n}]^*_{f-k}+ O(1/L). \EQN{5-apx6}
\eeqa
In \Ep{5-apx6}, $[\,\,]_{f-k}$ means that we exclude the $k$th field mode.
 With \Ep{5-apx6} and neglecting $O(1/L)$ terms, \Ep{1-apx6} becomes
\beqa
& &\int d\Gam \sum_k b_k J_k  [(e^{-i\tht t})^{\dagger}q_1^{m}q_1^{*n}]^* \rhot (\Gam,0)\EQN{6-apx6}  \\
& &= \sum_k \int d\Gam b_k J_k  [(e^{-i\tht t})^{\dagger}q_1^{m}q_1^{*n}]^*_{f-k} \rhot (\Gam,0) \nonumber \\
& &= \sum_k \int d\Gam b_k J_k  [(e^{-i\tht t})^{\dagger}q_1^{m}q_1^{*n}]^*_{f-k} g_1(\Gam_1)\prod_k g_k (\Gam_k) \nonumber \\
& &=  \sum_k \int d\Gam_k b_k J_k g_k (\Gam_k) \nonumber \\
& &\times \int d\Gam_{f-k}  [(e^{-i\tht
t})^{\dagger}q_1^{m}q_1^{*n}]^*_{f-k} g_1(\Gam_1)\prod_{k' \ne k}
g_{k'} (\Gam_{k'}). \nonumber   \eeqa %%
For any $k$ we have
\beqa
\int d\Gam_k g(\Gam_k) =1.
\eeqa
Then we can write
\beqa
& & \int d\Gam_k b_k J_k g_k (\Gam_k) \nonumber \\
& &=  \int d\Gam b_k J_k g_1(\Gam_1)\prod_k g_k (\Gam_k) = b_k \bra J_k \ket, \\
& & \int d\Gam_{f-k}  [(e^{-i\tht t})^{\dagger}q_1^{m}q_1^{*n}]^*_{f-k} g_1(\Gam_1)\prod_{k' \ne k} g_{k'} (\Gam_{k'}) \nonumber \\
& &= \int d\Gam  [(e^{-i\tht t})^{\dagger}q_1^{m}q_1^{*n}]^*_{f-k} g_1(\Gam_1)\prod_{k'} g_{k'} (\Gam_{k'}) \nonumber \\
& &= \int d\Gam q_1^{*m}q_1^n e^{-i\tht t} \rhot(\Gam,0),
\eeqa
and  \Ep{6-apx6} can be written as
\beqa
& &\int d\Gam \sum_k b_k J_k  [(e^{-i\tht t})^{\dagger}q_1^{m}q_1^{*n}]^* \rhot (\Gam,0) \nonumber \\
& &= \sum_k b_k \bra J_k \ket \int d\Gam q^{*m}_1 q_1^n e^{-i\tht t} \rhot (\Gam,0) \nonumber \\
& &=  \sum_k b_k \bra J_k \ket \int d\Gam q^{*m}_1 q_1^n \rhot (\Gam,t).
\eeqa
This equation, together with Eqs. (\ref{189-apx}) and (\ref{190-apx}), leads to \Ep{fac-sec-fp}.

%%%%%%%%%%%%%%%%%%%%
% References       %
%%%%%%%%%%%%%%%%%%%%

%\input fp-refs.tex

%%%%%%%%%%%%%%%%%%%%
% Figure Captions  %
%%%%%%%%%%%%%%%%%%%%

%\input captions.tex

\end{document}